\documentclass[twocolumn,superscriptaddress,prl,reprint,longbibliography
]{revtex4-1}
\usepackage{graphicx}
\usepackage{amssymb}
\usepackage{amsmath}
\usepackage{epsfig}
\usepackage{color}
\usepackage{natbib}
\usepackage{mathtools}
\usepackage[colorlinks,linkcolor=blue,anchorcolor=blue,citecolor=blue,urlcolor=blue]{hyperref}
\usepackage[left]{lineno}
\usepackage{blindtext}
\usepackage{color, soul}
\usepackage{setspace}
\usepackage{threeparttable}

\begin{document}

\title{Fast high-fidelity geometric quantum control with quantum brachistochrones}
\author{Yang Dong}
\author{Ce Feng}
\author{Yu Zheng}
\author{Xiang-Dong Chen}
\author{Guang-Can Guo}
\author{Fang-Wen Sun}
\email{fwsun@ustc.edu.cn}
\affiliation{{CAS Key Laboratory of Quantum Information, University of Science and Technology of China, Hefei, 230026, People's Republic of China}}
\affiliation{{CAS Center for Excellence in Quantum Information and Quantum Physics, University of Science and Technology of China, Hefei, 230026, People's Republic of China}}

\begin{abstract}
We experimentally demonstrate fast and high-fidelity geometric control of a quantum system with the most brachistochrone method on hybrid spin registers in diamond.
Based on the time-optimal universal geometric control, single geometric gates with the fidelities over $99.2{\%}$ on the spin state of nitrogen-vacancy center are realized with average durations shortened by $74.9{\%}$, comparing with conventional geometric method. The fidelity of the fast geometric two-qubit gate exceeds $96.5{\%}$ on the hybrid spin registers. With these fast high-fidelity gates available, we implement quantum entanglement-enhanced phase estimation algorithm and demonstrate the Heisenberg quantum limit at room-temperature. By comparing with the conventional geometric circuit, the measurement bandwidth and sensitivity is enhanced by $3.5$ and $2.9$ times. Hence, our results show that high-fidelity quantum control based on a fast geometric route will be a versatile tool for broad applications of quantum information processing in practice.
\end{abstract}
\date{\today}
\maketitle

Quantum information processing (QIP), which can provide an unprecedented supremacy \cite{arute2019quantum,alexeev2021quantum} over classical counterpart in searching algorithm \cite{abrams1999quantum,harrow2009quantum}, simulation \cite{buluta2009quantum,georgescu2014quantum}, metrology \cite{giovannetti2004quantum,giovannetti2006quantum,pirandola2018advances,pezze2018quantum,dong2016reviving,liu2015demonstration}, and secure communication \cite{xu2020secure}, relies seriously on high quality quantum operations \cite{rong2015experimental,dong2018non,dong2019robust,bourassa2020entanglement}.
Meanwhile, the geometric phase \cite{zu2014experimental,nagata2018universal,yuan2018observation} is currently a central topic not only in quantum physics but also in various scientific fields including optics, electronics, nanotechnology, chemical reactions, and materials science.
With the built-in noise-resilience feature, the geometric phase based holonomic quantum gate is believed to be an ideal way to build a fault-tolerant universal quantum computer \cite{duan2001geometric}. Based on adiabatic or nonadiabatic cyclical paths, it has been proposed as a promising method to realize high-fidelity and robust manipulation in theory \cite{duan2001geometric,sjoqvist2012non,xu2012nonadiabatic,liu2019plug,johansson2012robustness}, along with experimental demonstration in various physical platforms, including superconducting circuits \cite{abdumalikov2013experimental,yan2019experimental,xu2020experimental}, nuclear magnetic resonance \cite{feng2013experimental,zhu2019single}, and solid-state defect in diamond \cite{zu2014experimental,arroyo2014room,nagata2018universal,huang2019experimental,dong2021experimental}.

However, the duration times of these geometric operations are still much longer than dynamical cases, especially for adiabatic geometric quantum computation (GQC) protocol, resulting in more decoherence degrading quantum states over time \cite{vitanov2017stimulated,guery2019shortcuts,yan2019experimental,johansson2012robustness,zhu2019single,zheng2016comparison,li2020fast}. Besides fighting decoherence \cite{guery2019shortcuts,barry2020sensitivity,dong2021composite,dong2021quantifying}, long operation durations will severely reduce the detection sensitivity and bandwidth for quantum sensing based on full GQC.
%experiments need to be repeated more often to increase signal-to-noise ratios and enhance the detection bandwidth and sensitivity for quantum sensing based on full GQC.
Usually, time optimal control (TOC) method can provide a useful toolbox to accelerate operations. However, finding accurate TOC geometric operations is difficult because the closed evolution path, the optimization of fidelity and time should be satisfied simultaneously \cite{rezakhani2009quantum,chen2010shortcut,ban2012fast,daems2013robust,torosov2011high}.

Here, by employing quantum brachistochrone equation (QBE) \cite{wang2015quantum,carlini2006time,carlini2012time,geng2016experimental}, we implement fast and high-fidelity geometric quantum gates based on the TOC method. In the experiment, we demonstrate a single-loop brachistochrone non-adiabatic holonomic quantum computation (B-NHQC) in a prototype hybrid quantum system: a two-qubit register consisting of nitrogen-vacancy (NV) electron and a nuclear spin in diamond. Geometric single-qubit gates with fidelities over $99.2\%$ on electron spin are realized with the average cutoff duration shortened by $74.9\%$, comparing with conventional NHQC method.
The fast two-qubit geometric gate is also realized with a fidelity of $96.5\% $. Based on the fast and high fidelity single- and two-qubit gates, we further illustrate the power of the B-NHQC scheme by implementing the entanglement-enhanced phase estimation algorithm. In additional to the demonstration of the Heisenberg quantum limit (HQL) with the two entangled qubits \cite{giovannetti2004quantum,giovannetti2006quantum,pirandola2018advances,pezze2018quantum,dong2016reviving,liu2015demonstration}, the best detection sensitivity and bandwidth are further improved by $2.9$ and $3.5$ times, respectively, which benefits from fast geometric operation mainly. Our results, thus, suggest that B-NHQC based gates can provide a novel and promising route to achieve fast and precise universal quantum control and make an important step towards practical high performance application of QIP.

\begin{figure}[tbp]
\centering
\textsf{\includegraphics[width=8.5cm]{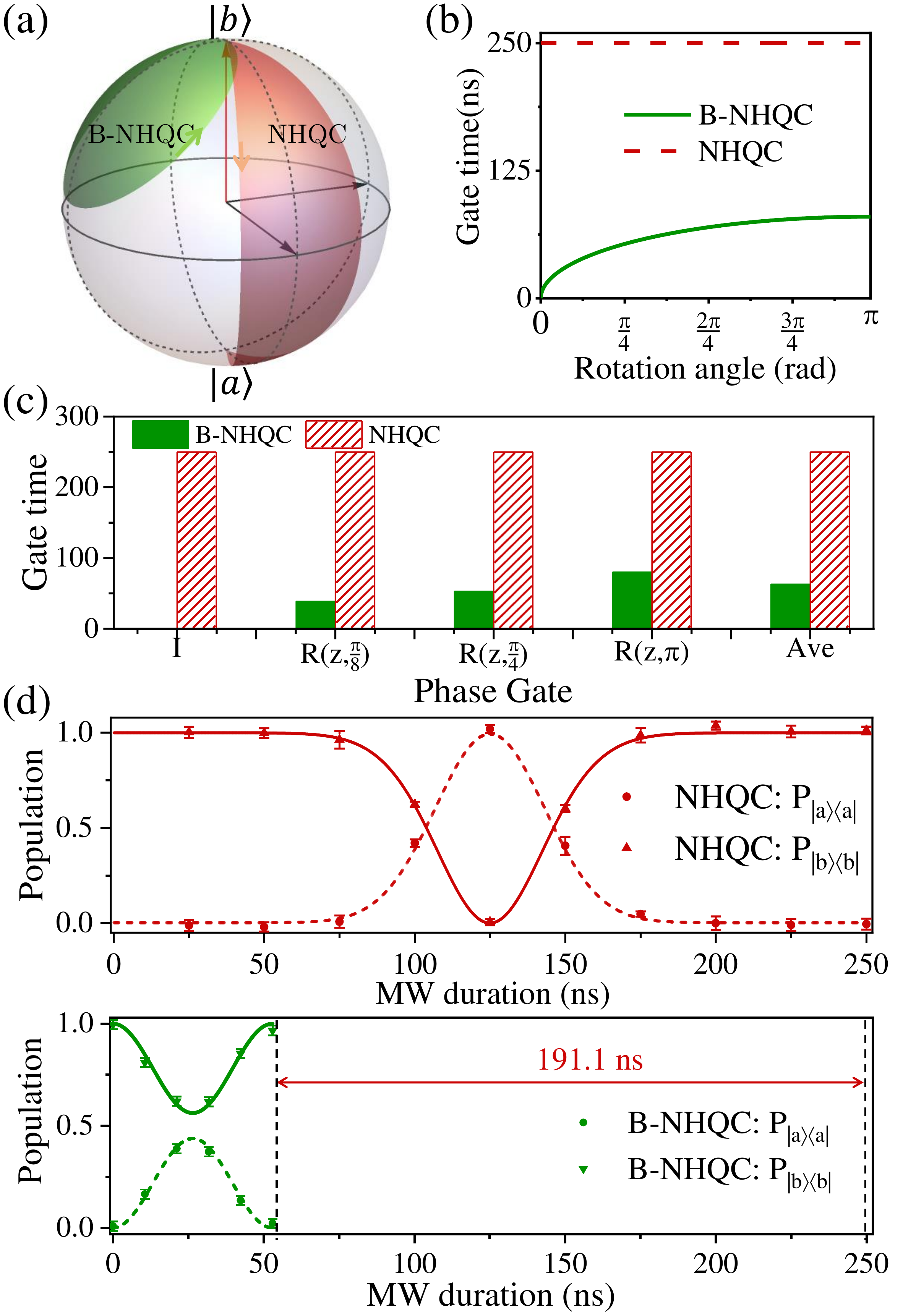}}
\caption{(a) A geometric picture of the quantum gates in the Bloch sphere representation shows the evolution paths of B-NHQC (green line) and conventional NHQC (red line) schemes. (b) Comparison of time costs for target gate operator
$R(\hat z,\theta )$ with $\theta  \in \left[ {0,\pi } \right]$. The maximum of the Rabi frequency of the Gaussian $4\sigma $ truncated pulse \cite{motzoi2009simple,abdumalikov2013experimental} is $12.76$ MHz for NHQC protocol \cite{SM}. (c) Comparison of the experimental gate time for $\theta  = \pi /8$, $\pi /4$, and $\pi $. (d) State evolutions during $T = R\left( {\hat z,\frac{\pi }{4}} \right)$ with B-NHQC and the conventional NHQC paths. The initial state is $\left| b \right\rangle  = \left( {\left| 0 \right\rangle  + \left| 1 \right\rangle } \right)/\sqrt 2 $. The time of the geometric gate with the TOC is $191.1$ ns ($75.4\%$) shorter than that of the conventional scheme. }
\label{fig1}
\end{figure}

We construct the geometric quantum gate \cite{arroyo2014room,zu2014experimental,zhu2019single,dong2021experimental} with the NV center in diamond \cite{maze2008nanoscale,xiang2013hybrid,dong2018non,lu2020observing}, a promising candidate for a scalable quantum register. By applying a magnetic field along the axis of the NV center, the electron spin-triplet states form a $V$-type energy level system (see Supplemental Material for details \cite{SM}). We encode $\left| {{m_s} =  - 1} \right\rangle  \equiv \left| 0 \right\rangle $ and
$\left| {{m_s} =  1} \right\rangle  \equiv \left| 1 \right\rangle $ as the qubit basis states, and use $\left| {{m_s} = 0} \right\rangle  \equiv \left| a \right\rangle $ as an ancillary state for the geometric manipulation of the qubit. Two microwave (MW) fields with frequencies ${\omega _{1(2)}}$ and initial phases ${\phi _{1(2)}}$ are used to couple the sequential transitions $\left| 0 \right\rangle  \leftrightarrow \left| a \right\rangle $ ($\left| a \right\rangle  \leftrightarrow \left| 1 \right\rangle $) with corresponding Rabi frequencies ${\Omega _{1(2)}}$, as described by \cite{SM}
\begin{eqnarray}
{H} =\frac{{\Omega (t)}}{2}{e^{i\phi _{2}}\left\vert b\right\rangle }\langle
a|+H.c. \text{,}  \label{EQ1}
\end{eqnarray}%
where ${\left\vert b\right\rangle ={e^{i\phi }}\sin \frac{\theta }{2}%
|0\rangle {+\cos}\frac{\theta }{2}|1\rangle }$ is a bright state with $\tan \frac{\theta }{2}=\frac{{\Omega
_{1}}}{{\Omega _{2}}}$.
Here, $\phi ={\phi _{1}}-{\phi _{2}}$ and $\Omega (t)=\sqrt{\Omega
_{1}^{2}(t)+\Omega _{2}^{2}(t)}$.
Therefore, the bright state ${\left\vert
b\right\rangle }$ interacts with the state $\left\vert a\right\rangle $, while decouples from the dark state ${\left| d \right\rangle  = \cos \frac{\theta }{2}|0\rangle  - {e^{ - i\phi }}\sin\frac{\theta }{2}|1\rangle }$.

The NHQC scheme \cite{dong2021experimental,SM} can be realized with a single-loop scenario by engineering the quantum system to evolve along with a red-slice-shaped path in the Bloch sphere, as shown in Fig. \ref{fig1}(a). In the qubit computational basis $\left\{ {|0\rangle ,|1\rangle } \right\}$, the gate operation is ${U_G}(\gamma ,\theta ,\phi ) = {e^{i(\gamma /2)}}{e^{ - i(\gamma /2)\vec n \cdot \vec \sigma }}$,
which describes a rotation around the $\vec n=(\sin \theta \cos \phi, \sin \theta \sin \phi, \cos \theta)$ axis by a $\gamma$ angle.
Therefore, the duration times are same for all gates, as shown in Figs. \ref{fig1}(b) and \ref{fig1}(c).
%And the smooth Gaussian $4\sigma $ truncated pulse is employed to eliminate unwanted population leakage effect during implementing the conventional NHQC protocol \cite{abdumalikov2013experimental,yan2019experimental,zu2014experimental,motzoi2009simple,johansson2012robustness}.

However, in realistic physical systems, the ever-present decoherence from the environment degrades the qualities of quantum states and operations over time. Therefore, generating the fastest possible cyclic path evolution by TOC becomes a preferable choice for realizing precise quantum controls \cite{ban2012fast,wang2015quantum,carlini2006time,carlini2012time}. And the accurate path with the minimal time cost can be obtained by solving the QBE together with the Schr\"{o}dinger equation. Here, the QBE can be written as $\dot{F}=-i[H, F]$, where $F=\partial L_{C} / \partial H$ and $L_{C}=\sum_{j} \lambda_{j} f_{j}(H)$ with the
Lagrange multiplier $\lambda_{j}$. One physically relevant constraint is the finite operation speed \cite{lloyd2000ultimate,lam2021demonstration} which is described as ${f_1}\left( {H(t)} \right) \equiv \frac{1}{2}\left[ {\operatorname{Tr} \left( {H{{(t)}^2}} \right) - \frac{{\Omega {{(t)}^2}}}{2}} \right] = 0,$ where ${\Omega (t)}$ is the Rabi frequency. The other constraint is ${f_2}\left( {H(t)} \right) \equiv Tr(H(t){\sigma _z}) = 0$ since we can not directly implement the independent ${\sigma _z}$ operation in the experiment.
A simple minimum-time solution \cite{liu2020brachistochronic,PhysRevApplied.14.034038} to this QBE is ${\phi _2}(t) = 2(\gamma  - \pi )t/\tau $, with the minimum gate duration time of $\tau_{min}= 2\sqrt {{\pi ^2} - {{(\pi  - \gamma )}^2}} /\Omega $, which increases with the geometric phase, as shown in Figs. \ref{fig1}(b) and \ref{fig1}(c). Under this condition, the single loop geometric gate $U_{G}$ can be implemented by a green-spherical-crown-shaped path as shown in Fig. \ref{fig1}(a), which is less time consuming than the conventional NHQC scheme as shown in red.

\begin{figure}[bp]
\centering
\textsf{\includegraphics[width=8.5cm]{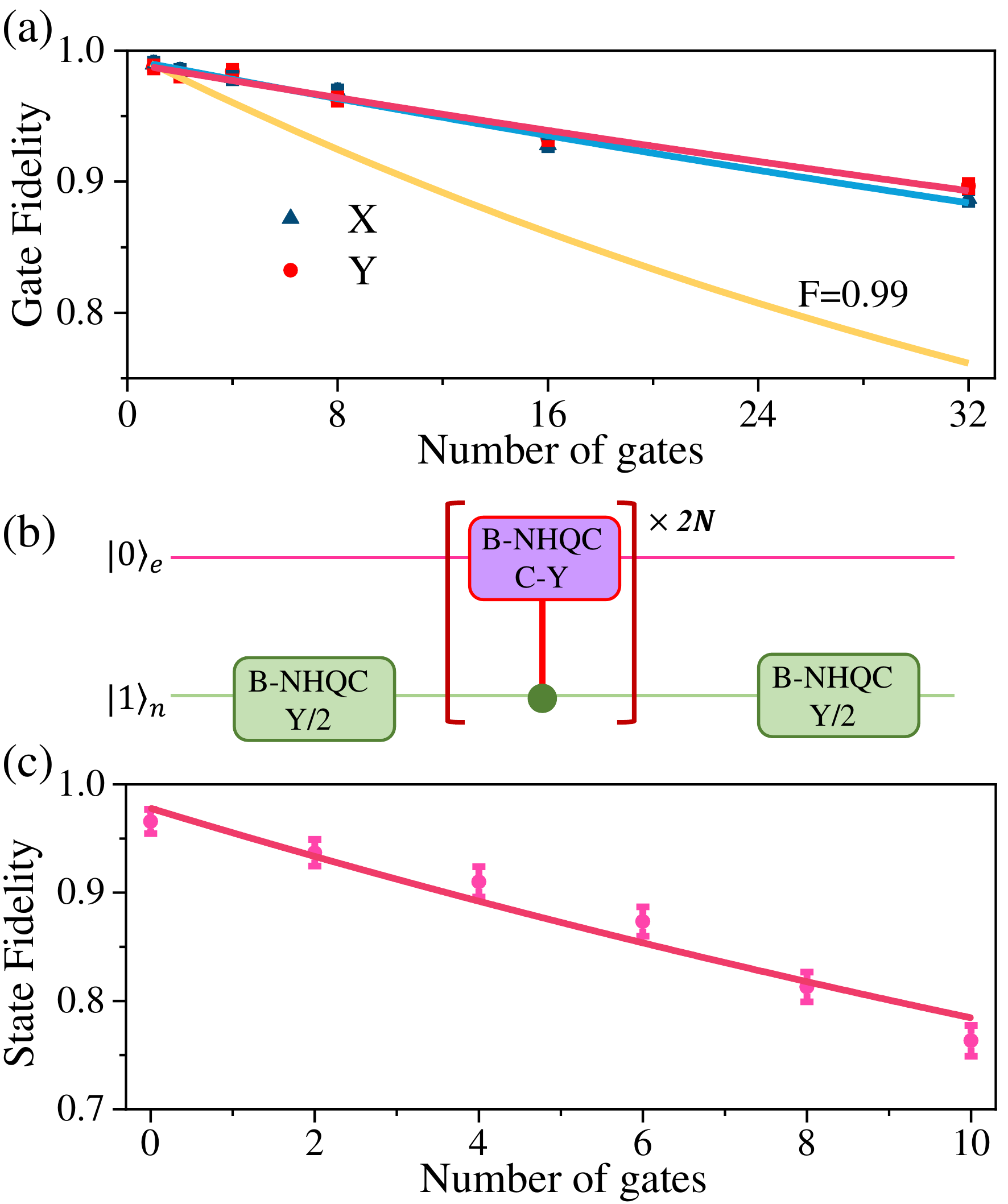}}
\caption{(a) The decay of the fidelity of the B-NHQC X,Y gates obtained via QPT. The golden line shows the decay with the gate fidelity of $F=0.9$. (b) The quantum circuit of $N$ B-NHQC C-Y gates. (c) Two-qubit state fidelities after $N$ applications of the C-Y operation.}
\label{fig2}
\end{figure}

By setting $\Omega (t) =12.5$ MHz, we experimentally demonstrate $R(\hat z,\theta )$ operation with $\theta  = \pi /8$, $\pi /4$, and $\pi $ based on the B-NHQC.
The time cost is significantly shorter than that with the NHQC scheme, on average, by $74.9\% $, as shown in Fig. \ref{fig1}(c). Especially, we implement the target $T = {U_G}\left( {\pi /4,0,0} \right)$ gate operation with both methods as shown in Fig. \ref{fig1}(d). The duration is only $24.6\%$ of that with the NHQC scheme.
%And the B-NHQC scheme can reduce the integrated other-state population compared with NHQC protocol, which are beneficial to reduce the transverse dephasing process of NV center.
We characterize the single-qubit B-NHQC gates through a standard quantum process tomography (QPT) method \cite{dong2021quantifying} on six specific geometric gates $\left\{ {I,X/2,X,Y/2,Y,T} \right\}$, with an average fidelity of 0.984(4) \cite{SM}. The major contribution to the QPT infidelity comes from state preparation and detection errors \cite{zu2014experimental,arroyo2014room,kleissler2018universal,huang2019experimental}. Therefore, we implement $X$ and $Y$ gates successively on the electron spins and perform QPT when the gates are repeated for $N$ times. The average fidelity \cite{zu2014experimental,rong2015experimental,rong2014implementation} of $N$ B-NHQC gates can be calculated as: $F_N=\left[1+\left(1-\varepsilon_{i f}\right)(1-p)^{N}\right] / 2$, where ${\varepsilon _{if}}$ describes errors in the state preparation and measurement and $p$ is the average error per gate \cite{zu2014experimental,rong2014implementation}, as shown in Fig. \ref{fig2} (a). We estimate the average fidelity of the $X$, $Y$ gates to be ${F_X} = 1 - p_X = 0.9922(4)$ and ${F_Y}= 0.9923(4)$ from the experimental data, which exceed the threshold of the necessary fidelity for the realization of the state-of-art error correction codes based on surface codes \cite{kleissler2018universal}.

\begin{table*}[btp]
\centering
\caption{Summary of the gate fidelity and duration for various QC schemes.}
\tabcolsep0.025in
\begin{tabular}{c c c c c c}
  \hline\hline
  Scheme & DD protected QC \cite{zhang2014protected,zhang2015experimental} & SUPCODE QC \cite{rong2014implementation}& GRAPE QC \cite{rong2015experimental} & AGQC \cite{huang2019experimental} & B-NHQC  \\
  \hline
  Single-bit fidelity & 0.98 & 0.9961(2) & 0.999952(6) & 0.991(2) & 0.9922(4)  \\
  Two-bit fidelity & 0.90 & -- & 0.9920(1) & 0.94(2) & 0.965(4)  \\
  Single-bit gate duration ($\mu$s) & 0.355   & 5.063  & 0.34  & 1  & 0.063  \\
  Two-bit gate duration ($\mu$s) & 50  & -- & 0.696  & 2  & 0.354  \\
  \hline\hline
\end{tabular}
\end{table*}

Universal geometric control of qubits always requires a nontrivial two-qubit gate.
Here, in the construction of two-qubit gate, we use the electron spin state as target state and the nitrogen nuclear spin ($^{14}\text{N}$) as the control qubit \cite{SM}.
%For this type of the solid hybrid spin system, the two-qubit gate poses some challenges in the experiment, mostly because the characteristic properties of the two types of the spins differ by three orders of magnitude.
In the experiment, we implement a controlled-rotation gate with MW selective operation, which is equivalent to a controlled-NOT gate \cite{jelezko2004observation,zhang2014protected,zhang2015experimental}. A typical controlled-Y (C-Y) gate can be represented as ${U_{\text{C-Y}}}=\left( {%
\begin{array}{cc}
\mathbf{1} & \mathbf{0} \\
\mathbf{0} & \mathbf{{{R_{y}}(\pi )}}%
\end{array}%
}\right) $,
where $\mathbf{1},\mathbf{0}$, and $\mathbf{{{R_y}(\pi )}}$ represent $2 \times 2$ matrices corresponding to the unit operator, zero matrix, and a rotation matrix around the $y$-axis with $\mathbf{R_y}\left( \pi  \right) = {e^{ - i\pi {\mathbf{\sigma}_y}/2}}$. Hence, the electron spin will be rotated if the $^{14}{\text{N}}$ nuclear spin state is $\left| {{m_I} = 1} \right\rangle $. We implement geometric $I$ gate if the $^{14}{\text{N}}$ nuclear spin state is $\left| {{m_I} = -1} \right\rangle $ with selective MW field driving instead of doing nothing. In this way, the coherence of this hybrid system can be protected by the dressed states \cite{golter2014protecting}.
%Since the inhomogeneous broadening effects of NV center's electron spin degrade the performance of the entangled gate, we implement geometric $I$ gate if the $^{14}{\text{N}}$ nuclear spin state is $\left| {{m_I} = -1} \right\rangle $ with selective MW field driving instead of doing nothing as shown in Fig. \ref{fig4}(b). In this way, the coherence of this hybrid system can be protected by the dressed states \cite{golter2014protecting}.
Using this method, the geometric entangled state between the electron and $^{14}{\text{N}}$ nuclear spin is prepared and measured in the Bell state $\left| \Phi  \right\rangle  = \left( {\left| {00} \right\rangle  + \left| {11} \right\rangle } \right)/\sqrt 2 $ with a state fidelity of ${F_s} = 0.947(8)$ by the quantum state tomography (QST) method \cite{SM}. By applying the B-NHQC C-Y gate repeatedly, the state of the hybrid system is coherently transferred among entangled and separable states, as shown in Fig. \ref{fig2}(b).
%In the hybrid system composed of the electron and nuclear spins, coherent operation of the nuclear spin costs much longer time than that of the electron spin. The typical operation time (${t_{\pi /2}} = 36.74$ $\mu {\text{s}}$) on the $^{14}{\text{N}}$ nuclear spin is much longer than the dephasing time ($T_2^* = 8.1(2)$ $\mu {\text{s}}$) of electron spin qubit \cite{zhang2015experimental}. The decoherence effect on the electron spin during the $^{14}{\text{N}}$ nuclear spin operation in the two-qubit gate experiments will dominate the fidelity decay. Hence, repeated applications of the C-Y gate \cite{chow2012universal} on the system and recording the dynamics of the quantum state is a better choice to estimate the gate fidelity as shown in Fig. \ref{fig4}(d).
Fig. \ref{fig2}(c) shows the fidelity of the final state to the ideal state for up to $10$ applications of the C-Y operation with no entanglement in the hybrid systems for even number of gates. The result reveals ${F_s(N)} = AF_g^N + B$, where $A$, $B$ and two-qubit gate fidelity ${F_g}$ are fit parameters \cite{chow2012universal}.
%When $N$ is increased, the error of the C-Y gate will accumulate and state fidelity will decay.
Here, we can obtain that the two-qubit gate fidelity is ${F_g} = 0.965(4)$.

In Table \uppercase\expandafter{\romannumeral1}, we make a comparison between various QC schemes with B-NHQC in experiment. The power of the MW or radio-frequency drives is almost the same for these schemes.
Since the B-NHQC scheme is the fastest geometric operation and the longest sequence duration is much less than the decoherence time, decoherence effects are expected to be negligible and the fluctuations of control fields will be the major error.
Fortunately, these control errors can be directly mitigated by combining B-NHQC with composite-pulse architecture \cite{zhu2019single,rong2014implementation,genov2017arbitrarily}.

%By employing composite-pulse idea \cite{zhu2019single,genov2017arbitrarily,zhang2014protected,zhang2015experimental,rong2014implementation,rong2015experimental}, the fidelity of the dynamical gate can be improved by consuming a significant amount of time.

%The shortest possible time duration of the B-NHQC is advantageous to high fidelity due to the reduction of the $^{14}{\text{N}}$ spin dephasing effect. As shown However, for B-NHQC scheme, high performance QIP can be realized with the least amount of time.

\begin{figure}[bp]
\centering
\textsf{\includegraphics[width=8.5cm]{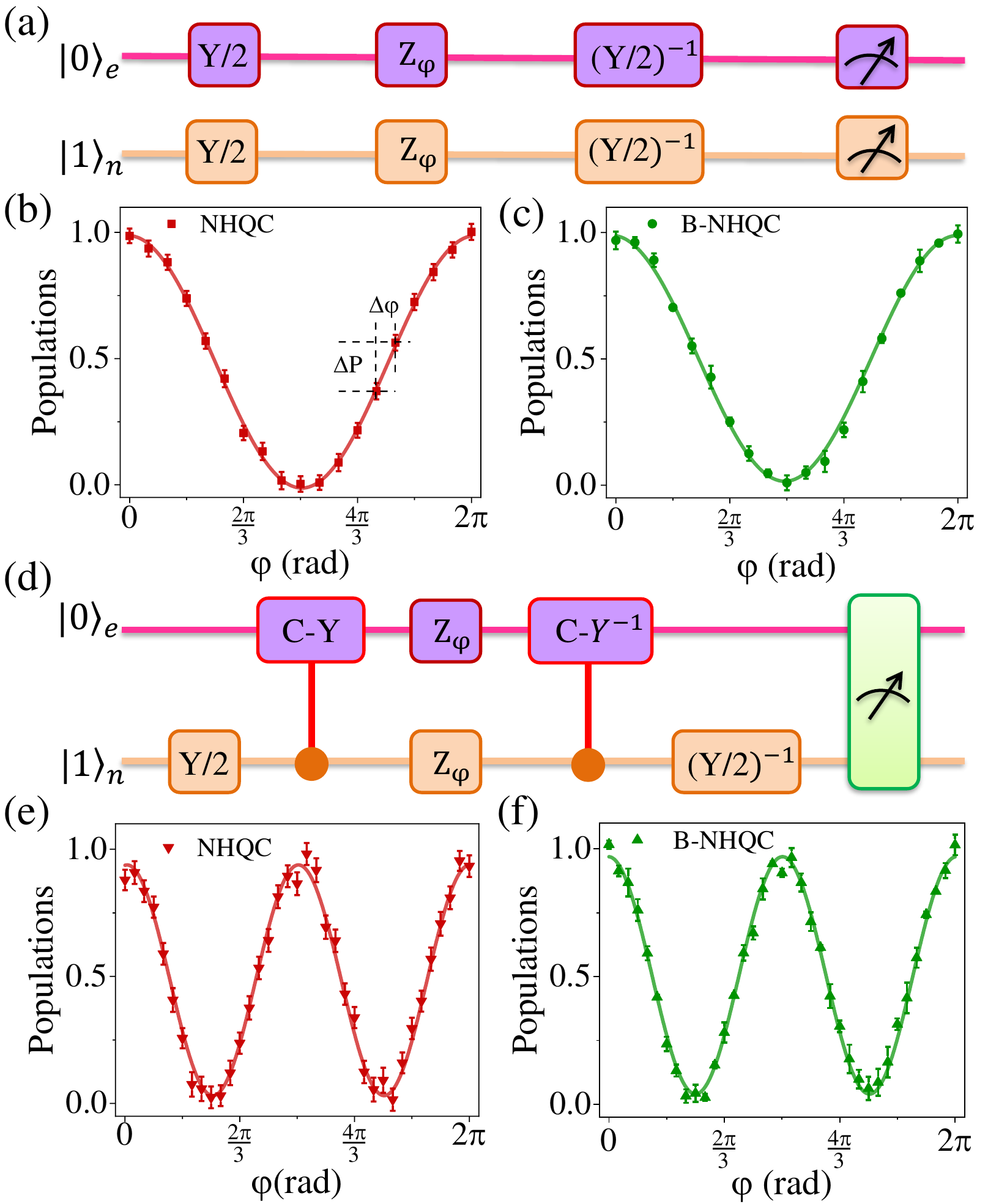}}
\caption{Demonstration of the HQL with entanglement-enhanced phase estimation algorithm with NHQC and B-NHQC gates. (a) Phase estimation schemes of the independent quantum probes. (b)-(c) Phase relation of single probe with NHQC and B-NHQC protocol. (d) Phase estimation schemes of two entangled quantum probes. (e)-(f) Phase relation of the two entangled probes with NHQC and B-NHQC gates. The phase gate ${Z_\varphi }$ is realized by adding a phase to
the drive field for all operations.}
\label{fig3}
\end{figure}

We further illustrate the advantage of QIP based on the full fast B-NHQC gates by implementing entanglement-enhanced parameter estimation algorithm \cite{giovannetti2004quantum,giovannetti2006quantum,pirandola2018advances,pezze2018quantum,dong2016reviving,liu2015demonstration} with the hybrid solid-state spin registers.
%Quantum-enhanced measurement will benefit all quantitative modern science and technology.
By using independent sensors to estimate a parameter, we can achieve at best the standard quantum limit. However, it is believed that using entangled quantum sensor qubits, such as NOON states \cite{giovannetti2004quantum,giovannetti2006quantum,pirandola2018advances,pezze2018quantum,dong2016reviving,liu2015demonstration}, are able to achieve the fundamental HQL scaling.
Fig. \ref{fig3} shows the quantum circuit to prepare quantum states and probe the phase $\varphi $ \cite{giovannetti2004quantum,giovannetti2006quantum,pirandola2018advances,pezze2018quantum,dong2016reviving,liu2015demonstration}, which can encode the quantity to be measured. For conventional phase estimation procedure in Fig. \ref{fig3}(a), the interferometer signal is the expectation value $P = ({{1 + \cos \varphi }})/{2}$ \cite{SM}, as shown in Figs. \ref{fig3}(b) and \ref{fig3}(c) with the photon-shot-noise at the output denoted by error bars. %If we repeat the same interferometric procedure with $2$ uncorrelated sensor qubits, the standard deviation of the photon-shot-noise will decrease by $\sqrt 2 $, showing the SQL.
Moreover, the electron and nuclear spin can be prepared in an entangled state. Shown in Fig. \ref{fig3}(d), we implement a phase gate on both sensor qubits, which brings the system to the two-qubit NOON state
$\left( {\left| {00} \right\rangle  + {e^{{-}2{\text{i}}\varphi }}\left| {11} \right\rangle } \right)/\sqrt 2 $. The interferometer signal is $P = ({{1 + \cos 2\varphi }})/{2}$, as shown in Figs. \ref{fig3}(e) and \ref{fig3}(f) with NHQC and B-NHQC protocol, respectively.
%The phase relation of the entangled state has double frequency dependence on the phase, which means the phase estimation employing the NOON state is more sensitive than that of two uncorrelated sensor qubits.

The quantum sensor is most sensitive to variations of the phase ($\Delta \varphi $) at the point of the maximum slope, with the sensitivity limited by the uncertainty in the interferometer signal measurement ($\Delta P$). The uncertainty of the hybrid spin register to small variations in phase, as depicted in the measurements, is given by $\Delta {\varphi _{\min }} = \sigma _S^n/dP$, where $\sigma _S^n$  is the standard deviation of the interferometer measurement after $n$ averages and $dP$ is the slope of the interferometer signal variation with phase \cite{giovannetti2004quantum,giovannetti2006quantum,pirandola2018advances,pezze2018quantum,maze2008nanoscale}.
By harnessing the entanglement of the electron and nuclear spins, the phase estimation uncertainties are reduced by factors of $1.93(1)$ for the NHQC and $1.99(1)$ for the B-NHQC protocol compared with single qubit, which is close to $2$ to demonstrate the HQL.
Because of reducing gate duration time significantly with the B-NHQC protocol, the results of the phase estimation is better than those from NHQC protocol, as shown in Figs. \ref{fig3}(c) and \ref{fig3}(f). The visibility \cite{SM} of the interferometer signal of the B-NHQC circuit is larger than the conventional scheme mainly by mitigating quantum decoherence effect with fast quantum control. Also, the instability of experimental environment is suppressed for fast quantum circuit. As a result, the fluctuation ($\sigma _S^n$) of interferometer signal is reduced in the experiment, which is denoted with error bars in Fig. \ref{fig3}. Here, average error bars levels are $\sigma _{S,NHQC}^n = 0.044(4)$ and $\sigma _{S,B - NHQC}^n = 0.031(3)$.

Moreover, to comprehensively and systematically analyze the experimental results, the duration time of both protocols should be taken into consideration to discuss the measurement sensitivity \cite{maze2008nanoscale,barry2020sensitivity,dong2021quantifying}. The best measurement sensitivity is given by
$S  = \Delta {\varphi _{\min }}\sqrt  T_t$,
where $T_t  $ is the total measurement time.
%To quantitatively compare entanglement enhanced phase estimation based on the B-NHQC or NHQC protocol, we list the measured parameters in Table \uppercase\expandafter{\romannumeral2}.
The phase estimation sensitivity enhancement is a direct and effective indicator, which can be calculated by
\begin{eqnarray}
\kappa  = \frac{{{S_{NHQC}}}}{{{S_{B - NHQC}}}}{\text{  }} = \frac{{{{\left( {\sigma _S^n{{\sqrt T }_t}/dP} \right)}_{NHQC}}}}{{{{\left( {\sigma _S^n{{\sqrt T }_t}/dP} \right)}_{B - NHQC}}}}\text{.}
\label{EQ6}
\end{eqnarray}%
Here, the slope ($dP$) is proportional to the visibility. The total single measurement time for the B-NHQC circuit includes the optical initial time (${T_{ini}} = 3$ $\mu$s), the duration of entanglement-enhanced phase estimation algorithm (${T_a} = 2T\left( {{^{14}{\text{N}}},\frac{\pi }{2}} \right) + 2T\left( {e,\pi } \right) \approx 79.1$ $\mu$s), and quantum state readout time ${T_r} = 2$ $\mu$s. So the sensing time is mostly spent on the entanglement-enhanced phase estimation algorithm, which is more prominent for NHQC circuit (${T_a} \approx 287.0$ $\mu$s). Therefore, the sensitivity enhancement is $\kappa  = 2.9$, which benefits from fast geometric operation mainly (${T_{t,NHQC}}/{T_{t,B - NHQC}} = 3.5$). Because of the reduction in time cost, the signal detection bandwidth is improved by $3.5$ times with the B-NHQC circuit.

In summary, fast and high-fidelity quantum coherent control is of fundamental significance in quantum computation, simulation, and metrology. We have implemented a robust B-NHQC based single- and two-qubit gate with hybrid spin registers in diamond, which paves the way for full fast geometric QIP in practice.
%The average gate duration is reduced by $74.9\%$ compared with original NHQC scheme. Based on the B-NHQC scheme, a programmable solid-state quantum information processor can be constructed efficiently at room-temperature. Moreover, by employing geometric single- and two-qubit gates, we also perform a quantum entanglement-enhanced phase estimation algorithm. And benefiting from short quantum control time, the measurement sensitivity and bandwidth is improved significantly, which is compatible with high-resolution NMR detection at the nanoscale and far beyond the canonical entanglement-enhanced phase estimation paradigm.
And benefiting from short quantum control time, the measurement sensitivity and bandwidth in the entanglement-enhanced phase estimation is improved significantly, which is far beyond the conventional estimation paradigm. It will be further applied in high-resolution nuclear magnetic resonance detection at the nanoscale \cite{zaiser2016enhancing,aslam2017nanoscale,barry2020sensitivity,dong2021quantifying}. Beyond the demonstration in our work, the B-NHQC protocol can be applied to other promising physical platforms, such as atoms, ions, transmon or flux qubits \cite{xiang2013hybrid}, for fast and high-fidelity QIP.

We thank Zheng-Yuan Xue for valuable discussions. This work is supported by the National Key Research and Development Program of China (Grant No. 2017YFA0304504), the National Natural Science Foundation of China (Grants No. 91850102 and No. 12005218), the Anhui Initiative in Quantum Information Technologies (Grant No. AHY130000), the Science Challenge Project (Grant No. TZ2018003), the Fundamental Research Funds for the Central Universities (Grant No. WK2030000020) and Excellent Youth Foundation of Hebei Scientific Committee (Grant No. F2019516002).

%\bibliography{myrefdate}

\begin{thebibliography}{67}
\expandafter\ifx\csname natexlab\endcsname\relax\def\natexlab#1{#1}\fi
\expandafter\ifx\csname bibnamefont\endcsname\relax
  \def\bibnamefont#1{#1}\fi
\expandafter\ifx\csname bibfnamefont\endcsname\relax
  \def\bibfnamefont#1{#1}\fi
\expandafter\ifx\csname citenamefont\endcsname\relax
  \def\citenamefont#1{#1}\fi
\expandafter\ifx\csname url\endcsname\relax
  \def\url#1{\texttt{#1}}\fi
\expandafter\ifx\csname urlprefix\endcsname\relax\def\urlprefix{URL }\fi
\providecommand{\bibinfo}[2]{#2}
\providecommand{\eprint}[2][]{\url{#2}}

\bibitem[{\citenamefont{Arute et~al.}(2019)\citenamefont{Arute, Arya, Babbush,
  Bacon, Bardin, Barends, Biswas, Boixo, Brandao, Buell
  et~al.}}]{arute2019quantum}
\bibinfo{author}{\bibfnamefont{F.}~\bibnamefont{Arute}},
  \bibinfo{author}{\bibfnamefont{K.}~\bibnamefont{Arya}},
  \bibinfo{author}{\bibfnamefont{R.}~\bibnamefont{Babbush}},
  \bibinfo{author}{\bibfnamefont{D.}~\bibnamefont{Bacon}},
  \bibinfo{author}{\bibfnamefont{J.~C.} \bibnamefont{Bardin}},
  \bibinfo{author}{\bibfnamefont{R.}~\bibnamefont{Barends}},
  \bibinfo{author}{\bibfnamefont{R.}~\bibnamefont{Biswas}},
  \bibinfo{author}{\bibfnamefont{S.}~\bibnamefont{Boixo}},
  \bibinfo{author}{\bibfnamefont{F.~G.} \bibnamefont{Brandao}},
  \bibinfo{author}{\bibfnamefont{D.~A.} \bibnamefont{Buell}},
  \bibnamefont{et~al.}, \emph{\bibinfo{title}{Quantum Supremacy Using a Programmable Superconducting Processor}}, \bibinfo{journal}{Nature}
  \textbf{\bibinfo{volume}{574}}, \bibinfo{pages}{505} (\bibinfo{year}{2019}).

\bibitem[{\citenamefont{Alexeev et~al.}(2021)\citenamefont{Alexeev, Bacon,
  Brown, Calderbank, Carr, Chong, DeMarco, Englund, Farhi, Fefferman
  et~al.}}]{alexeev2021quantum}
\bibinfo{author}{\bibfnamefont{Y.}~\bibnamefont{Alexeev}},
  \bibinfo{author}{\bibfnamefont{D.}~\bibnamefont{Bacon}},
  \bibinfo{author}{\bibfnamefont{K.~R.} \bibnamefont{Brown}},
  \bibinfo{author}{\bibfnamefont{R.}~\bibnamefont{Calderbank}},
  \bibinfo{author}{\bibfnamefont{L.~D.} \bibnamefont{Carr}},
  \bibinfo{author}{\bibfnamefont{F.~T.} \bibnamefont{Chong}},
  \bibinfo{author}{\bibfnamefont{B.}~\bibnamefont{DeMarco}},
  \bibinfo{author}{\bibfnamefont{D.}~\bibnamefont{Englund}},
  \bibinfo{author}{\bibfnamefont{E.}~\bibnamefont{Farhi}},
  \bibinfo{author}{\bibfnamefont{B.}~\bibnamefont{Fefferman}},
  \bibnamefont{et~al.}, \emph{\bibinfo{title}{Quantum Computer Systems for Scientific Discovery}}, \bibinfo{journal}{PRX Quantum}
  \textbf{\bibinfo{volume}{2}}, \bibinfo{pages}{017001} (\bibinfo{year}{2021}).

\bibitem[{\citenamefont{Abrams and Lloyd}(1999)}]{abrams1999quantum}
\bibinfo{author}{\bibfnamefont{D.~S.} \bibnamefont{Abrams}} \bibnamefont{and}
  \bibinfo{author}{\bibfnamefont{S.}~\bibnamefont{Lloyd}},
  \emph{\bibinfo{title}{Quantum Algorithm Providing Exponential Speed Increase for Finding Eigenvalues and Eigenvectors}},
  \bibinfo{journal}{Phys. Rev. Lett.} \textbf{\bibinfo{volume}{83}},
  \bibinfo{pages}{5162} (\bibinfo{year}{1999}).

\bibitem[{\citenamefont{Harrow et~al.}(2009)\citenamefont{Harrow, Hassidim, and
  Lloyd}}]{harrow2009quantum}
\bibinfo{author}{\bibfnamefont{A.~W.} \bibnamefont{Harrow}},
  \bibinfo{author}{\bibfnamefont{A.}~\bibnamefont{Hassidim}}, \bibnamefont{and}
  \bibinfo{author}{\bibfnamefont{S.}~\bibnamefont{Lloyd}},
  \emph{\bibinfo{title}{Quantum Algorithm for Linear Systems of Equations}},
  \bibinfo{journal}{Phys. Rev. Lett.} \textbf{\bibinfo{volume}{103}},
  \bibinfo{pages}{150502} (\bibinfo{year}{2009}).

\bibitem[{\citenamefont{Buluta and Nori}(2009)}]{buluta2009quantum}
\bibinfo{author}{\bibfnamefont{I.}~\bibnamefont{Buluta}} \bibnamefont{and}
  \bibinfo{author}{\bibfnamefont{F.}~\bibnamefont{Nori}},
  \emph{\bibinfo{title}{Quantum Simulators}},
  \bibinfo{journal}{Science} \textbf{\bibinfo{volume}{326}},
  \bibinfo{pages}{108} (\bibinfo{year}{2009}).

\bibitem[{\citenamefont{Georgescu et~al.}(2014)\citenamefont{Georgescu, Ashhab,
  and Nori}}]{georgescu2014quantum}
\bibinfo{author}{\bibfnamefont{I.~M.} \bibnamefont{Georgescu}},
  \bibinfo{author}{\bibfnamefont{S.}~\bibnamefont{Ashhab}}, \bibnamefont{and}
  \bibinfo{author}{\bibfnamefont{F.}~\bibnamefont{Nori}},
  \emph{\bibinfo{title}{Quantum Simulation}}, \bibinfo{journal}{Rev. Mod.
  Phys.} \textbf{\bibinfo{volume}{86}}, \bibinfo{pages}{153}
  (\bibinfo{year}{2014}).

\bibitem[{\citenamefont{Giovannetti et~al.}(2004)\citenamefont{Giovannetti,
  Lloyd, and Maccone}}]{giovannetti2004quantum}
\bibinfo{author}{\bibfnamefont{V.}~\bibnamefont{Giovannetti}},
  \bibinfo{author}{\bibfnamefont{S.}~\bibnamefont{Lloyd}}, \bibnamefont{and}
  \bibinfo{author}{\bibfnamefont{L.}~\bibnamefont{Maccone}},
  \emph{\bibinfo{title}{Quantum-Enhanced Measurements: Beating the Standard Quantum Limit}},
  \bibinfo{journal}{Science} \textbf{\bibinfo{volume}{306}},
  \bibinfo{pages}{1330} (\bibinfo{year}{2004}).

\bibitem[{\citenamefont{Giovannetti et~al.}(2006)\citenamefont{Giovannetti,
  Lloyd, and Maccone}}]{giovannetti2006quantum}
\bibinfo{author}{\bibfnamefont{V.}~\bibnamefont{Giovannetti}},
  \bibinfo{author}{\bibfnamefont{S.}~\bibnamefont{Lloyd}}, \bibnamefont{and}
  \bibinfo{author}{\bibfnamefont{L.}~\bibnamefont{Maccone}},
  \emph{\bibinfo{title}{Quantum Metrology}},
  \bibinfo{journal}{Phys. Rev. Lett.} \textbf{\bibinfo{volume}{96}},
  \bibinfo{pages}{010401} (\bibinfo{year}{2006}).

\bibitem[{\citenamefont{Pirandola et~al.}(2018)\citenamefont{Pirandola,
  Bardhan, Gehring, Weedbrook, and Lloyd}}]{pirandola2018advances}
\bibinfo{author}{\bibfnamefont{S.}~\bibnamefont{Pirandola}},
  \bibinfo{author}{\bibfnamefont{B.~R.} \bibnamefont{Bardhan}},
  \bibinfo{author}{\bibfnamefont{T.}~\bibnamefont{Gehring}},
  \bibinfo{author}{\bibfnamefont{C.}~\bibnamefont{Weedbrook}},
  \bibnamefont{and} \bibinfo{author}{\bibfnamefont{S.}~\bibnamefont{Lloyd}},
  \emph{\bibinfo{title}{Advances in Photonic Quantum Sensing}},
  \bibinfo{journal}{Nat. Photonics} \textbf{\bibinfo{volume}{12}},
  \bibinfo{pages}{724} (\bibinfo{year}{2018}).

\bibitem[{\citenamefont{Pezze et~al.}(2018)\citenamefont{Pezze, Smerzi,
  Oberthaler, Schmied, and Treutlein}}]{pezze2018quantum}
\bibinfo{author}{\bibfnamefont{L.}~\bibnamefont{Pezze}},
  \bibinfo{author}{\bibfnamefont{A.}~\bibnamefont{Smerzi}},
  \bibinfo{author}{\bibfnamefont{M.~K.} \bibnamefont{Oberthaler}},
  \bibinfo{author}{\bibfnamefont{R.}~\bibnamefont{Schmied}}, \bibnamefont{and}
  \bibinfo{author}{\bibfnamefont{P.}~\bibnamefont{Treutlein}},
  \emph{\bibinfo{title}{Quantum Metrology with Nonclassical States of Atomic Ensembles}}, \bibinfo{journal}{Rev. Mod. Phys.}
  \textbf{\bibinfo{volume}{90}}, \bibinfo{pages}{035005}
  (\bibinfo{year}{2018}).

\bibitem[{\citenamefont{Dong et~al.}(2016)\citenamefont{Dong, Chen, Guo, and
  Sun}}]{dong2016reviving}
\bibinfo{author}{\bibfnamefont{Y.}~\bibnamefont{Dong}},
  \bibinfo{author}{\bibfnamefont{X.-D.} \bibnamefont{Chen}},
  \bibinfo{author}{\bibfnamefont{G.-C.} \bibnamefont{Guo}}, \bibnamefont{and}
  \bibinfo{author}{\bibfnamefont{F.-W.} \bibnamefont{Sun}},
  \emph{\bibinfo{title}{Reviving the Precision of Multiple Entangled Probes in an Open System by Simple $\pi$-Pulse Sequences}}, \bibinfo{journal}{Phys.
  Rev. A} \textbf{\bibinfo{volume}{94}}, \bibinfo{pages}{052322}
  (\bibinfo{year}{2016}).

\bibitem[{\citenamefont{Liu et~al.}(2015)\citenamefont{Liu, Zhang, Chang, Yue,
  Fan, and Pan}}]{liu2015demonstration}
\bibinfo{author}{\bibfnamefont{G.-Q.} \bibnamefont{Liu}},
  \bibinfo{author}{\bibfnamefont{Y.-R.} \bibnamefont{Zhang}},
  \bibinfo{author}{\bibfnamefont{Y.-C.} \bibnamefont{Chang}},
  \bibinfo{author}{\bibfnamefont{J.-D.} \bibnamefont{Yue}},
  \bibinfo{author}{\bibfnamefont{H.}~\bibnamefont{Fan}}, \bibnamefont{and}
  \bibinfo{author}{\bibfnamefont{X.-Y.} \bibnamefont{Pan}},
  \emph{\bibinfo{title}{Demonstration of Entanglement-Enhanced Phase Estimation in Solid}}, \bibinfo{journal}{Nat. Commun.} \textbf{\bibinfo{volume}{6}},
  \bibinfo{pages}{1} (\bibinfo{year}{2015}).

\bibitem[{\citenamefont{Xu et~al.}(2020{\natexlab{a}})\citenamefont{Xu, Ma,
  Zhang, Lo, and Pan}}]{xu2020secure}
\bibinfo{author}{\bibfnamefont{F.}~\bibnamefont{Xu}},
  \bibinfo{author}{\bibfnamefont{X.}~\bibnamefont{Ma}},
  \bibinfo{author}{\bibfnamefont{Q.}~\bibnamefont{Zhang}},
  \bibinfo{author}{\bibfnamefont{H.-K.} \bibnamefont{Lo}}, \bibnamefont{and}
  \bibinfo{author}{\bibfnamefont{J.-W.} \bibnamefont{Pan}},
  \emph{\bibinfo{title}{Secure Quantum Key Distribution with Realistic Devices}}, \bibinfo{journal}{Rev. Mod. Phys.} \textbf{\bibinfo{volume}{92}},
  \bibinfo{pages}{025002} (\bibinfo{year}{2020}{\natexlab{a}}).

\bibitem[{\citenamefont{Rong et~al.}(2015)\citenamefont{Rong, Geng, Shi, Liu,
  Xu, Ma, Kong, Jiang, Wu, and Du}}]{rong2015experimental}
\bibinfo{author}{\bibfnamefont{X.}~\bibnamefont{Rong}},
  \bibinfo{author}{\bibfnamefont{J.}~\bibnamefont{Geng}},
  \bibinfo{author}{\bibfnamefont{F.}~\bibnamefont{Shi}},
  \bibinfo{author}{\bibfnamefont{Y.}~\bibnamefont{Liu}},
  \bibinfo{author}{\bibfnamefont{K.}~\bibnamefont{Xu}},
  \bibinfo{author}{\bibfnamefont{W.}~\bibnamefont{Ma}},
  \bibinfo{author}{\bibfnamefont{F.}~\bibnamefont{Kong}},
  \bibinfo{author}{\bibfnamefont{Z.}~\bibnamefont{Jiang}},
  \bibinfo{author}{\bibfnamefont{Y.}~\bibnamefont{Wu}}, \bibnamefont{and}
  \bibinfo{author}{\bibfnamefont{J.}~\bibnamefont{Du}},
  \emph{\bibinfo{title}{Experimental Fault-Tolerant Universal Quantum Gates with Solid-State Spins under Ambient Conditions}}, \bibinfo{journal}{Nat.
  Commun.} \textbf{\bibinfo{volume}{6}}, \bibinfo{pages}{1}
  (\bibinfo{year}{2015}).

\bibitem[{\citenamefont{Dong et~al.}(2018{\natexlab{a}})\citenamefont{Dong,
  Zheng, Li, Li, Chen, Guo, and Sun}}]{dong2018non}
\bibinfo{author}{\bibfnamefont{Y.}~\bibnamefont{Dong}},
  \bibinfo{author}{\bibfnamefont{Y.}~\bibnamefont{Zheng}},
  \bibinfo{author}{\bibfnamefont{S.}~\bibnamefont{Li}},
  \bibinfo{author}{\bibfnamefont{C.-C.} \bibnamefont{Li}},
  \bibinfo{author}{\bibfnamefont{X.-D.} \bibnamefont{Chen}},
  \bibinfo{author}{\bibfnamefont{G.-C.} \bibnamefont{Guo}}, \bibnamefont{and}
  \bibinfo{author}{\bibfnamefont{F.-W.} \bibnamefont{Sun}},
  \emph{\bibinfo{title}{Non-Markovianity-Assisted High-Fidelity Deutsch-Jozsa Algorithm in Diamond}}, \bibinfo{journal}{NPJ Quantum Inf.}
  \textbf{\bibinfo{volume}{4}}, \bibinfo{pages}{1}
  (\bibinfo{year}{2018}{\natexlab{a}}).

\bibitem[{\citenamefont{Dong et~al.}(2019)\citenamefont{Dong, Chen, Guo, and
  Sun}}]{dong2019robust}
\bibinfo{author}{\bibfnamefont{Y.}~\bibnamefont{Dong}},
  \bibinfo{author}{\bibfnamefont{X.-D.} \bibnamefont{Chen}},
  \bibinfo{author}{\bibfnamefont{G.-C.} \bibnamefont{Guo}}, \bibnamefont{and}
  \bibinfo{author}{\bibfnamefont{F.-W.} \bibnamefont{Sun}},
  \emph{\bibinfo{title}{Robust Scalable Architecture for a Hybrid Spin-Mechanical Quantum Entanglement System}}, \bibinfo{journal}{Phys. Rev.
  B} \textbf{\bibinfo{volume}{100}}, \bibinfo{pages}{214103}
  (\bibinfo{year}{2019}).

\bibitem[{\citenamefont{Bourassa et~al.}(2020)\citenamefont{Bourassa, Anderson,
  Miao, Onizhuk, Ma, Crook, Abe, Ul-Hassan, Ohshima, Son
  et~al.}}]{bourassa2020entanglement}
\bibinfo{author}{\bibfnamefont{A.}~\bibnamefont{Bourassa}},
  \bibinfo{author}{\bibfnamefont{C.~P.} \bibnamefont{Anderson}},
  \bibinfo{author}{\bibfnamefont{K.~C.} \bibnamefont{Miao}},
  \bibinfo{author}{\bibfnamefont{M.}~\bibnamefont{Onizhuk}},
  \bibinfo{author}{\bibfnamefont{H.}~\bibnamefont{Ma}},
  \bibinfo{author}{\bibfnamefont{A.~L.} \bibnamefont{Crook}},
  \bibinfo{author}{\bibfnamefont{H.}~\bibnamefont{Abe}},
  \bibinfo{author}{\bibfnamefont{J.}~\bibnamefont{Ul-Hassan}},
  \bibinfo{author}{\bibfnamefont{T.}~\bibnamefont{Ohshima}},
  \bibinfo{author}{\bibfnamefont{N.~T.} \bibnamefont{Son}},
  \bibnamefont{et~al.}, \emph{\bibinfo{title}{Entanglement and Control of Single Nuclear Spins in Isotopically Engineered Silicon Carbide}},
  \bibinfo{journal}{Nat. Mater.} \textbf{\bibinfo{volume}{19}},
  \bibinfo{pages}{1319} (\bibinfo{year}{2020}).

\bibitem[{\citenamefont{Zu et~al.}(2014)\citenamefont{Zu, Wang, He, Zhang, Dai,
  Wang, and Duan}}]{zu2014experimental}
\bibinfo{author}{\bibfnamefont{C.}~\bibnamefont{Zu}},
  \bibinfo{author}{\bibfnamefont{W.-B.} \bibnamefont{Wang}},
  \bibinfo{author}{\bibfnamefont{L.}~\bibnamefont{He}},
  \bibinfo{author}{\bibfnamefont{W.-G.} \bibnamefont{Zhang}},
  \bibinfo{author}{\bibfnamefont{C.-Y.} \bibnamefont{Dai}},
  \bibinfo{author}{\bibfnamefont{F.}~\bibnamefont{Wang}}, \bibnamefont{and}
  \bibinfo{author}{\bibfnamefont{L.-M.} \bibnamefont{Duan}},
  \emph{\bibinfo{title}{Experimental Realization of Universal Geometric Quantum Gates with Solid-State Spins}}, \bibinfo{journal}{Nature}
  \textbf{\bibinfo{volume}{514}}, \bibinfo{pages}{72} (\bibinfo{year}{2014}).

\bibitem[{\citenamefont{Nagata et~al.}(2018)\citenamefont{Nagata, Kuramitani,
  Sekiguchi, and Kosaka}}]{nagata2018universal}
\bibinfo{author}{\bibfnamefont{K.}~\bibnamefont{Nagata}},
  \bibinfo{author}{\bibfnamefont{K.}~\bibnamefont{Kuramitani}},
  \bibinfo{author}{\bibfnamefont{Y.}~\bibnamefont{Sekiguchi}},
  \bibnamefont{and} \bibinfo{author}{\bibfnamefont{H.}~\bibnamefont{Kosaka}},
  \emph{\bibinfo{title}{Universal Holonomic Quantum Gates over Geometric Spin Qubits with Polarised Microwaves}}, \bibinfo{journal}{Nat. Commun.}
  \textbf{\bibinfo{volume}{9}}, \bibinfo{pages}{1} (\bibinfo{year}{2018}).

\bibitem[{\citenamefont{Yuan et~al.}(2018)\citenamefont{Yuan, Guan, Chen, Zhao,
  Yu, Luo, Tan, Xie, Wang, Sun et~al.}}]{yuan2018observation}
\bibinfo{author}{\bibfnamefont{D.}~\bibnamefont{Yuan}},
  \bibinfo{author}{\bibfnamefont{Y.}~\bibnamefont{Guan}},
  \bibinfo{author}{\bibfnamefont{W.}~\bibnamefont{Chen}},
  \bibinfo{author}{\bibfnamefont{H.}~\bibnamefont{Zhao}},
  \bibinfo{author}{\bibfnamefont{S.}~\bibnamefont{Yu}},
  \bibinfo{author}{\bibfnamefont{C.}~\bibnamefont{Luo}},
  \bibinfo{author}{\bibfnamefont{Y.}~\bibnamefont{Tan}},
  \bibinfo{author}{\bibfnamefont{T.}~\bibnamefont{Xie}},
  \bibinfo{author}{\bibfnamefont{X.}~\bibnamefont{Wang}},
  \bibinfo{author}{\bibfnamefont{Z.}~\bibnamefont{Sun}}, \bibnamefont{et~al.},
  \emph{\bibinfo{title}{Observation of the Geometric Phase Effect in the
  $h+hd \to {H_2}+ d$ Reaction}}, \bibinfo{journal}{Science}
  \textbf{\bibinfo{volume}{362}}, \bibinfo{pages}{1289} (\bibinfo{year}{2018}).

\bibitem[{\citenamefont{Duan et~al.}(2001)\citenamefont{Duan, Cirac, and
  Zoller}}]{duan2001geometric}
\bibinfo{author}{\bibfnamefont{L.-M.} \bibnamefont{Duan}},
  \bibinfo{author}{\bibfnamefont{J.~I.} \bibnamefont{Cirac}}, \bibnamefont{and}
  \bibinfo{author}{\bibfnamefont{P.}~\bibnamefont{Zoller}},
  \emph{\bibinfo{title}{Geometric Manipulation of Trapped Ions for Quantum Computation}}, \bibinfo{journal}{Science} \textbf{\bibinfo{volume}{292}},
  \bibinfo{pages}{1695} (\bibinfo{year}{2001}).

\bibitem[{\citenamefont{Sj{\"o}qvist et~al.}(2012)\citenamefont{Sj{\"o}qvist,
  Tong, Andersson, Hessmo, Johansson, and Singh}}]{sjoqvist2012non}
\bibinfo{author}{\bibfnamefont{E.}~\bibnamefont{Sj{\"o}qvist}},
  \bibinfo{author}{\bibfnamefont{D.-M.} \bibnamefont{Tong}},
  \bibinfo{author}{\bibfnamefont{L.~M.} \bibnamefont{Andersson}},
  \bibinfo{author}{\bibfnamefont{B.}~\bibnamefont{Hessmo}},
  \bibinfo{author}{\bibfnamefont{M.}~\bibnamefont{Johansson}},
  \bibnamefont{and} \bibinfo{author}{\bibfnamefont{K.}~\bibnamefont{Singh}},
  \emph{\bibinfo{title}{Non-Adiabatic Holonomic Quantum Computation}},
  \bibinfo{journal}{New J. Phys.} \textbf{\bibinfo{volume}{14}},
  \bibinfo{pages}{103035} (\bibinfo{year}{2012}).

\bibitem[{\citenamefont{Xu et~al.}(2012)\citenamefont{Xu, Zhang, Tong,
  Sj{\"o}qvist, and Kwek}}]{xu2012nonadiabatic}
\bibinfo{author}{\bibfnamefont{G.}~\bibnamefont{Xu}},
  \bibinfo{author}{\bibfnamefont{J.}~\bibnamefont{Zhang}},
  \bibinfo{author}{\bibfnamefont{D.}~\bibnamefont{Tong}},
  \bibinfo{author}{\bibfnamefont{E.}~\bibnamefont{Sj{\"o}qvist}},
  \bibnamefont{and} \bibinfo{author}{\bibfnamefont{L.}~\bibnamefont{Kwek}},
  \emph{\bibinfo{title}{Nonadiabatic Holonomic Quantum Computation in Decoherence-Free Subspaces}}, \bibinfo{journal}{Phys. Rev. Lett.}
  \textbf{\bibinfo{volume}{109}}, \bibinfo{pages}{170501}
  (\bibinfo{year}{2012}).

\bibitem[{\citenamefont{Liu et~al.}(2019)\citenamefont{Liu, Song, Xue, Wang,
  and Yung}}]{liu2019plug}
\bibinfo{author}{\bibfnamefont{B.-J.} \bibnamefont{Liu}},
  \bibinfo{author}{\bibfnamefont{X.-K.} \bibnamefont{Song}},
  \bibinfo{author}{\bibfnamefont{Z.-Y.} \bibnamefont{Xue}},
  \bibinfo{author}{\bibfnamefont{X.}~\bibnamefont{Wang}}, \bibnamefont{and}
  \bibinfo{author}{\bibfnamefont{M.-H.} \bibnamefont{Yung}},
  \emph{\bibinfo{title}{Plug-and-Play Approach to Nonadiabatic Geometric Quantum Gates}}, \bibinfo{journal}{Phys. Rev. Lett.}
  \textbf{\bibinfo{volume}{123}}, \bibinfo{pages}{100501}
  (\bibinfo{year}{2019}).

\bibitem[{\citenamefont{Johansson et~al.}(2012)\citenamefont{Johansson,
  Sj{\"o}qvist, Andersson, Ericsson, Hessmo, Singh, and
  Tong}}]{johansson2012robustness}
\bibinfo{author}{\bibfnamefont{M.}~\bibnamefont{Johansson}},
  \bibinfo{author}{\bibfnamefont{E.}~\bibnamefont{Sj{\"o}qvist}},
  \bibinfo{author}{\bibfnamefont{L.~M.} \bibnamefont{Andersson}},
  \bibinfo{author}{\bibfnamefont{M.}~\bibnamefont{Ericsson}},
  \bibinfo{author}{\bibfnamefont{B.}~\bibnamefont{Hessmo}},
  \bibinfo{author}{\bibfnamefont{K.}~\bibnamefont{Singh}}, \bibnamefont{and}
  \bibinfo{author}{\bibfnamefont{D.}~\bibnamefont{Tong}},
  \emph{\bibinfo{title}{Robustness of Nonadiabatic Holonomic Gates}},
  \bibinfo{journal}{Phys. Rev. A} \textbf{\bibinfo{volume}{86}},
  \bibinfo{pages}{062322} (\bibinfo{year}{2012}).

\bibitem[{\citenamefont{Abdumalikov~Jr
  et~al.}(2013)\citenamefont{Abdumalikov~Jr, Fink, Juliusson, Pechal, Berger,
  Wallraff, and Filipp}}]{abdumalikov2013experimental}
\bibinfo{author}{\bibfnamefont{A.~A.} \bibnamefont{Abdumalikov~Jr}},
  \bibinfo{author}{\bibfnamefont{J.~M.} \bibnamefont{Fink}},
  \bibinfo{author}{\bibfnamefont{K.}~\bibnamefont{Juliusson}},
  \bibinfo{author}{\bibfnamefont{M.}~\bibnamefont{Pechal}},
  \bibinfo{author}{\bibfnamefont{S.}~\bibnamefont{Berger}},
  \bibinfo{author}{\bibfnamefont{A.}~\bibnamefont{Wallraff}}, \bibnamefont{and}
  \bibinfo{author}{\bibfnamefont{S.}~\bibnamefont{Filipp}},
  \emph{\bibinfo{title}{Experimental realization of non-Abelian non-adiabatic geometric gates}},
  \bibinfo{journal}{Nature} \textbf{\bibinfo{volume}{496}},
  \bibinfo{pages}{482} (\bibinfo{year}{2013}).


\bibitem[{\citenamefont{Yan et~al.}(2019)\citenamefont{Yan, Liu, Xu, Song, Liu,
  Zhang, Deng, Yan, Rong, Huang et~al.}}]{yan2019experimental}
\bibinfo{author}{\bibfnamefont{T.}~\bibnamefont{Yan}},
  \bibinfo{author}{\bibfnamefont{B.-J.} \bibnamefont{Liu}},
  \bibinfo{author}{\bibfnamefont{K.}~\bibnamefont{Xu}},
  \bibinfo{author}{\bibfnamefont{C.}~\bibnamefont{Song}},
  \bibinfo{author}{\bibfnamefont{S.}~\bibnamefont{Liu}},
  \bibinfo{author}{\bibfnamefont{Z.}~\bibnamefont{Zhang}},
  \bibinfo{author}{\bibfnamefont{H.}~\bibnamefont{Deng}},
  \bibinfo{author}{\bibfnamefont{Z.}~\bibnamefont{Yan}},
  \bibinfo{author}{\bibfnamefont{H.}~\bibnamefont{Rong}},
  \bibinfo{author}{\bibfnamefont{K.}~\bibnamefont{Huang}},
  \bibnamefont{et~al.}, \emph{\bibinfo{title}{Experimental Realization of Nonadiabatic Shortcut to Non-Abelian Geometric Gates}},
  \bibinfo{journal}{Phys. Rev. Lett.} \textbf{\bibinfo{volume}{122}},
  \bibinfo{pages}{080501} (\bibinfo{year}{2019}).

\bibitem[{\citenamefont{Xu et~al.}(2020{\natexlab{b}})\citenamefont{Xu, Hua,
  Chen, Pan, Li, Han, Cai, Ma, Wang, Song et~al.}}]{xu2020experimental}
\bibinfo{author}{\bibfnamefont{Y.}~\bibnamefont{Xu}},
  \bibinfo{author}{\bibfnamefont{Z.}~\bibnamefont{Hua}},
  \bibinfo{author}{\bibfnamefont{T.}~\bibnamefont{Chen}},
  \bibinfo{author}{\bibfnamefont{X.}~\bibnamefont{Pan}},
  \bibinfo{author}{\bibfnamefont{X.}~\bibnamefont{Li}},
  \bibinfo{author}{\bibfnamefont{J.}~\bibnamefont{Han}},
  \bibinfo{author}{\bibfnamefont{W.}~\bibnamefont{Cai}},
  \bibinfo{author}{\bibfnamefont{Y.}~\bibnamefont{Ma}},
  \bibinfo{author}{\bibfnamefont{H.}~\bibnamefont{Wang}},
  \bibinfo{author}{\bibfnamefont{Y.}~\bibnamefont{Song}}, \bibnamefont{et~al.},
  \emph{\bibinfo{title}{Experimental Implementation of Universal Nonadiabatic Geometric Quantum Gates in a Superconducting Circuit}},
  \bibinfo{journal}{Phys. Rev. Lett.} \textbf{\bibinfo{volume}{124}},
  \bibinfo{pages}{230503} (\bibinfo{year}{2020}{\natexlab{b}}).

\bibitem[{\citenamefont{Feng et~al.}(2013)\citenamefont{Feng, Xu, and
  Long}}]{feng2013experimental}
\bibinfo{author}{\bibfnamefont{G.}~\bibnamefont{Feng}},
  \bibinfo{author}{\bibfnamefont{G.}~\bibnamefont{Xu}}, \bibnamefont{and}
  \bibinfo{author}{\bibfnamefont{G.}~\bibnamefont{Long}},
  \emph{\bibinfo{title}{Experimental Realization of Nonadiabatic Holonomic Quantum Computation}}, \bibinfo{journal}{Phys. Rev. Lett.}
  \textbf{\bibinfo{volume}{110}}, \bibinfo{pages}{190501}
  (\bibinfo{year}{2013}).

\bibitem[{\citenamefont{Zhu et~al.}(2019)\citenamefont{Zhu, Chen, Yang, Bian,
  Xue, and Peng}}]{zhu2019single}
\bibinfo{author}{\bibfnamefont{Z.}~\bibnamefont{Zhu}},
  \bibinfo{author}{\bibfnamefont{T.}~\bibnamefont{Chen}},
  \bibinfo{author}{\bibfnamefont{X.}~\bibnamefont{Yang}},
  \bibinfo{author}{\bibfnamefont{J.}~\bibnamefont{Bian}},
  \bibinfo{author}{\bibfnamefont{Z.-Y.} \bibnamefont{Xue}}, \bibnamefont{and}
  \bibinfo{author}{\bibfnamefont{X.}~\bibnamefont{Peng}},
  \emph{\bibinfo{title}{Single-Loop and Composite-Loop Realization of Nonadiabatic Holonomic Quantum Gates in a Decoherence-Free Subspace}},
  \bibinfo{journal}{Phys. Rev. Appl.} \textbf{\bibinfo{volume}{12}},
  \bibinfo{pages}{024024} (\bibinfo{year}{2019}).

\bibitem[{\citenamefont{Arroyo-Camejo et~al.}(2014)\citenamefont{Arroyo-Camejo,
  Lazariev, Hell, and Balasubramanian}}]{arroyo2014room}
\bibinfo{author}{\bibfnamefont{S.}~\bibnamefont{Arroyo-Camejo}},
  \bibinfo{author}{\bibfnamefont{A.}~\bibnamefont{Lazariev}},
  \bibinfo{author}{\bibfnamefont{S.~W.} \bibnamefont{Hell}}, \bibnamefont{and}
  \bibinfo{author}{\bibfnamefont{G.}~\bibnamefont{Balasubramanian}},
  \emph{\bibinfo{title}{Room Temperature High-Fidelity Holonomic Single-Qubit Gate on a Solid-State Spin}}, \bibinfo{journal}{Nat. Commun.}
  \textbf{\bibinfo{volume}{5}}, \bibinfo{pages}{4870} (\bibinfo{year}{2014}).

\bibitem[{\citenamefont{Huang et~al.}(2019)\citenamefont{Huang, Wu, Wang, Hou,
  Wang, Zhang, Lian, Liu, Wang, Zhang et~al.}}]{huang2019experimental}
\bibinfo{author}{\bibfnamefont{Y.-Y.} \bibnamefont{Huang}},
  \bibinfo{author}{\bibfnamefont{Y.-K.} \bibnamefont{Wu}},
  \bibinfo{author}{\bibfnamefont{F.}~\bibnamefont{Wang}},
  \bibinfo{author}{\bibfnamefont{P.-Y.} \bibnamefont{Hou}},
  \bibinfo{author}{\bibfnamefont{W.-B.} \bibnamefont{Wang}},
  \bibinfo{author}{\bibfnamefont{W.-G.} \bibnamefont{Zhang}},
  \bibinfo{author}{\bibfnamefont{W.-Q.} \bibnamefont{Lian}},
  \bibinfo{author}{\bibfnamefont{Y.-Q.} \bibnamefont{Liu}},
  \bibinfo{author}{\bibfnamefont{H.-Y.} \bibnamefont{Wang}},
  \bibinfo{author}{\bibfnamefont{H.-Y.} \bibnamefont{Zhang}},
  \bibnamefont{et~al.}, \emph{\bibinfo{title}{Experimental Realization of Robust Geometric Quantum Gates with Solid-State Spins}},
  \bibinfo{journal}{Phys. Rev. Lett.} \textbf{\bibinfo{volume}{122}},
  \bibinfo{pages}{010503} (\bibinfo{year}{2019}).

  \bibitem[{\citenamefont{Dong et~al.}(2021)\citenamefont{Dong,
  Zhang, Zheng, Lin, Shan, Chen, Zhu, Wang, Guo, and Sun}}]{dong2021experimental}
\bibinfo{author}{\bibfnamefont{Y.}~\bibnamefont{Dong}},
  \bibinfo{author}{\bibfnamefont{S.-C.} \bibnamefont{Zhang}},
  \bibinfo{author}{\bibfnamefont{Y.}~\bibnamefont{Zheng}},
  \bibinfo{author}{\bibfnamefont{H.-B.} \bibnamefont{Lin}},
  \bibinfo{author}{\bibfnamefont{L.-K.} \bibnamefont{Shan}},
  \bibinfo{author}{\bibfnamefont{X.-D.} \bibnamefont{Chen}},
  \bibinfo{author}{\bibfnamefont{W.}~\bibnamefont{Zhu}},
  \bibinfo{author}{\bibfnamefont{G.-Z.} \bibnamefont{Wang}},
  \bibinfo{author}{\bibfnamefont{G.-C.} \bibnamefont{Guo}}, \bibnamefont{and}
  \bibinfo{author}{\bibfnamefont{F.-W.} \bibnamefont{Sun}},
  \emph{\bibinfo{title}{Experimental implementation of universal holonomic quantum computation on solid-state spins with optimal control}},
  \bibinfo{journal}{arXiv:2102.09227}  (\bibinfo{year}{2021}).

\bibitem[{\citenamefont{Vitanov et~al.}(2017)\citenamefont{Vitanov, Rangelov,
  Shore, and Bergmann}}]{vitanov2017stimulated}
\bibinfo{author}{\bibfnamefont{N.~V.} \bibnamefont{Vitanov}},
  \bibinfo{author}{\bibfnamefont{A.~A.} \bibnamefont{Rangelov}},
  \bibinfo{author}{\bibfnamefont{B.~W.} \bibnamefont{Shore}}, \bibnamefont{and}
  \bibinfo{author}{\bibfnamefont{K.}~\bibnamefont{Bergmann}},
  \emph{\bibinfo{title}{Stimulated Raman Adiabatic Passage in Physics, Chemistry, and Beyond}}, \bibinfo{journal}{Rev. Mod. Phys.}
  \textbf{\bibinfo{volume}{89}}, \bibinfo{pages}{015006}
  (\bibinfo{year}{2017}).

\bibitem[{\citenamefont{Gu{\'e}ry-Odelin
  et~al.}(2019)\citenamefont{Gu{\'e}ry-Odelin, Ruschhaupt, Kiely, Torrontegui,
  Mart{\'\i}nez-Garaot, and Muga}}]{guery2019shortcuts}
\bibinfo{author}{\bibfnamefont{D.}~\bibnamefont{Gu{\'e}ry-Odelin}},
  \bibinfo{author}{\bibfnamefont{A.}~\bibnamefont{Ruschhaupt}},
  \bibinfo{author}{\bibfnamefont{A.}~\bibnamefont{Kiely}},
  \bibinfo{author}{\bibfnamefont{E.}~\bibnamefont{Torrontegui}},
  \bibinfo{author}{\bibfnamefont{S.}~\bibnamefont{Mart{\'\i}nez-Garaot}},
  \bibnamefont{and} \bibinfo{author}{\bibfnamefont{J.~G.} \bibnamefont{Muga}},
  \emph{\bibinfo{title}{Shortcuts to Adiabaticity: Concepts, Methods, and Applications}}, \bibinfo{journal}{Rev. Mod. Phys.}
  \textbf{\bibinfo{volume}{91}}, \bibinfo{pages}{045001}
  (\bibinfo{year}{2019}).

\bibitem[{\citenamefont{Zheng et~al.}(2016)\citenamefont{Zheng, Yang, and
  Nori}}]{zheng2016comparison}
\bibinfo{author}{\bibfnamefont{S.-B.} \bibnamefont{Zheng}},
  \bibinfo{author}{\bibfnamefont{C.-P.} \bibnamefont{Yang}}, \bibnamefont{and}
  \bibinfo{author}{\bibfnamefont{F.}~\bibnamefont{Nori}},
  \emph{\bibinfo{title}{Comparison of the Sensitivity to Systematic Errors between Nonadiabatic Non-Abelian Geometric Gates and Their Dynamical Counterparts}}, \bibinfo{journal}{Phys. Rev. A}
  \textbf{\bibinfo{volume}{93}}, \bibinfo{pages}{032313}
  (\bibinfo{year}{2016}).

\bibitem[{\citenamefont{Li et~al.}(2020)\citenamefont{Li, Chen, and
  Xue}}]{li2020fast}
\bibinfo{author}{\bibfnamefont{S.}~\bibnamefont{Li}},
  \bibinfo{author}{\bibfnamefont{T.}~\bibnamefont{Chen}}, \bibnamefont{and}
  \bibinfo{author}{\bibfnamefont{Z.-Y.} \bibnamefont{Xue}},
  \emph{\bibinfo{title}{Fast Holonomic Quantum Computation on Superconducting Circuits with Optimal Control}}, \bibinfo{journal}{Adv. Quantum Technol.}
  \textbf{\bibinfo{volume}{3}}, \bibinfo{pages}{2000001}
  (\bibinfo{year}{2020}).

\bibitem[{\citenamefont{Barry et~al.}(2020)\citenamefont{Barry, Schloss, Bauch,
  Turner, Hart, Pham, and Walsworth}}]{barry2020sensitivity}
\bibinfo{author}{\bibfnamefont{J.~F.} \bibnamefont{Barry}},
  \bibinfo{author}{\bibfnamefont{J.~M.} \bibnamefont{Schloss}},
  \bibinfo{author}{\bibfnamefont{E.}~\bibnamefont{Bauch}},
  \bibinfo{author}{\bibfnamefont{M.~J.} \bibnamefont{Turner}},
  \bibinfo{author}{\bibfnamefont{C.~A.} \bibnamefont{Hart}},
  \bibinfo{author}{\bibfnamefont{L.~M.} \bibnamefont{Pham}}, \bibnamefont{and}
  \bibinfo{author}{\bibfnamefont{R.~L.} \bibnamefont{Walsworth}},
  \emph{\bibinfo{title}{Sensitivity Optimization for NV-Diamond Magnetometry}},
  \bibinfo{journal}{Rev. Mod. Phys.} \textbf{\bibinfo{volume}{92}},
  \bibinfo{pages}{015004} (\bibinfo{year}{2020}).

\bibitem[{\citenamefont{Dong et~al.}(2021)\citenamefont{Dong, Zhang, Lin, Chen,
  Zhu, Wang, Guo, and Sun}}]{dong2021quantifying}
\bibinfo{author}{\bibfnamefont{Y.}~\bibnamefont{Dong}},
  \bibinfo{author}{\bibfnamefont{S.-C.} \bibnamefont{Zhang}},
  \bibinfo{author}{\bibfnamefont{H.-B.} \bibnamefont{Lin}},
  \bibinfo{author}{\bibfnamefont{X.-D.} \bibnamefont{Chen}},
  \bibinfo{author}{\bibfnamefont{W.}~\bibnamefont{Zhu}},
  \bibinfo{author}{\bibfnamefont{G.-Z.} \bibnamefont{Wang}},
  \bibinfo{author}{\bibfnamefont{G.-C.} \bibnamefont{Guo}}, \bibnamefont{and}
  \bibinfo{author}{\bibfnamefont{F.-W.} \bibnamefont{Sun}},
  \emph{\bibinfo{title}{Quantifying the Performance of Multipulse Quantum Sensing}},
  \bibinfo{journal}{Phys. Rev. B} \textbf{\bibinfo{volume}{103}},
  \bibinfo{pages}{104104} (\bibinfo{year}{2021}).

\bibitem[{\citenamefont{Dong et~al.}(2021)\citenamefont{Dong, Xu,
  Zhang, Zheng, Chen, Zhu, Wang, Guo, and Sun}}]{dong2021composite}
\bibinfo{author}{\bibfnamefont{Y.}~\bibnamefont{Dong}},
  \bibinfo{author}{\bibfnamefont{J.-Y.} \bibnamefont{Xu}},
  \bibinfo{author}{\bibfnamefont{S.-C.} \bibnamefont{Zhang}},
  \bibinfo{author}{\bibfnamefont{Y.}~\bibnamefont{Zheng}},
  \bibinfo{author}{\bibfnamefont{X.-D.} \bibnamefont{Chen}},
  \bibinfo{author}{\bibfnamefont{W.}~\bibnamefont{Zhu}},
  \bibinfo{author}{\bibfnamefont{G.-Z.} \bibnamefont{Wang}},
  \bibinfo{author}{\bibfnamefont{G.-C.} \bibnamefont{Guo}}, \bibnamefont{and}
  \bibinfo{author}{\bibfnamefont{F.-W.} \bibnamefont{Sun}},
  \emph{\bibinfo{title}{Composite-pulse enhanced room-temperature diamond magnetometry}},
  \bibinfo{journal}{Funct. Diam.}
  \textbf{\bibinfo{volume}{1}}, \bibinfo{pages}{125} (\bibinfo{year}{2021}).

\bibitem[{\citenamefont{Rezakhani et~al.}(2009)\citenamefont{Rezakhani, Kuo,
  Hamma, Lidar, and Zanardi}}]{rezakhani2009quantum}
\bibinfo{author}{\bibfnamefont{A.}~\bibnamefont{Rezakhani}},
  \bibinfo{author}{\bibfnamefont{W.-J.} \bibnamefont{Kuo}},
  \bibinfo{author}{\bibfnamefont{A.}~\bibnamefont{Hamma}},
  \bibinfo{author}{\bibfnamefont{D.}~\bibnamefont{Lidar}}, \bibnamefont{and}
  \bibinfo{author}{\bibfnamefont{P.}~\bibnamefont{Zanardi}},
  \emph{\bibinfo{title}{Quantum Adiabatic Brachistochrone}},
  \bibinfo{journal}{Phys. Rev. Lett.} \textbf{\bibinfo{volume}{103}},
  \bibinfo{pages}{080502} (\bibinfo{year}{2009}).

\bibitem[{\citenamefont{Chen et~al.}(2010)\citenamefont{Chen, Lizuain,
  Ruschhaupt, Gu{\'e}ry-Odelin, and Muga}}]{chen2010shortcut}
\bibinfo{author}{\bibfnamefont{X.}~\bibnamefont{Chen}},
  \bibinfo{author}{\bibfnamefont{I.}~\bibnamefont{Lizuain}},
  \bibinfo{author}{\bibfnamefont{A.}~\bibnamefont{Ruschhaupt}},
  \bibinfo{author}{\bibfnamefont{D.}~\bibnamefont{Gu{\'e}ry-Odelin}},
  \bibnamefont{and} \bibinfo{author}{\bibfnamefont{J.}~\bibnamefont{Muga}},
  \emph{\bibinfo{title}{Shortcut to Adiabatic Passage in Two- and Three-Level Atoms}}, \bibinfo{journal}{Phys. Rev. Lett.} \textbf{\bibinfo{volume}{105}},
  \bibinfo{pages}{123003} (\bibinfo{year}{2010}).

\bibitem[{\citenamefont{Ban et~al.}(2012)\citenamefont{Ban, Chen, Sherman, and
  Muga}}]{ban2012fast}
\bibinfo{author}{\bibfnamefont{Y.}~\bibnamefont{Ban}},
  \bibinfo{author}{\bibfnamefont{X.}~\bibnamefont{Chen}},
  \bibinfo{author}{\bibfnamefont{E.~Y.} \bibnamefont{Sherman}},
  \bibnamefont{and} \bibinfo{author}{\bibfnamefont{J.}~\bibnamefont{Muga}},
  \emph{\bibinfo{title}{Fast and Robust Spin Manipulation in a Quantum Dot by Electric Fields}}, \bibinfo{journal}{Phys. Rev. Lett.}
  \textbf{\bibinfo{volume}{109}}, \bibinfo{pages}{206602}
  (\bibinfo{year}{2012}).

\bibitem[{\citenamefont{Daems et~al.}(2013)\citenamefont{Daems, Ruschhaupt,
  Sugny, and Guerin}}]{daems2013robust}
\bibinfo{author}{\bibfnamefont{D.}~\bibnamefont{Daems}},
  \bibinfo{author}{\bibfnamefont{A.}~\bibnamefont{Ruschhaupt}},
  \bibinfo{author}{\bibfnamefont{D.}~\bibnamefont{Sugny}}, \bibnamefont{and}
  \bibinfo{author}{\bibfnamefont{S.}~\bibnamefont{Guerin}},
  \emph{\bibinfo{title}{Robust Quantum Control by a Single-Shot Shaped Pulse}},
  \bibinfo{journal}{Phys. Rev. Lett.} \textbf{\bibinfo{volume}{111}},
  \bibinfo{pages}{050404} (\bibinfo{year}{2013}).

\bibitem[{\citenamefont{Torosov et~al.}(2011)\citenamefont{Torosov, Gu{\'e}rin,
  and Vitanov}}]{torosov2011high}
\bibinfo{author}{\bibfnamefont{B.~T.} \bibnamefont{Torosov}},
  \bibinfo{author}{\bibfnamefont{S.}~\bibnamefont{Gu{\'e}rin}},
  \bibnamefont{and} \bibinfo{author}{\bibfnamefont{N.~V.}
  \bibnamefont{Vitanov}}, \emph{\bibinfo{title}{High-Fidelity Adiabatic Passage by Composite Sequences of Chirped Pulses}}, \bibinfo{journal}{Phys. Rev.
  Lett.} \textbf{\bibinfo{volume}{106}}, \bibinfo{pages}{233001}
  (\bibinfo{year}{2011}).

\bibitem[{\citenamefont{Wang et~al.}(2015)\citenamefont{Wang, Allegra, Jacobs,
  Lloyd, Lupo, and Mohseni}}]{wang2015quantum}
\bibinfo{author}{\bibfnamefont{X.}~\bibnamefont{Wang}},
  \bibinfo{author}{\bibfnamefont{M.}~\bibnamefont{Allegra}},
  \bibinfo{author}{\bibfnamefont{K.}~\bibnamefont{Jacobs}},
  \bibinfo{author}{\bibfnamefont{S.}~\bibnamefont{Lloyd}},
  \bibinfo{author}{\bibfnamefont{C.}~\bibnamefont{Lupo}}, \bibnamefont{and}
  \bibinfo{author}{\bibfnamefont{M.}~\bibnamefont{Mohseni}},
  \emph{\bibinfo{title}{Quantum Brachistochrone Curves as Geodesics: Obtaining Accurate Minimum-Time Protocols for the Control of Quantum Systems}},
  \bibinfo{journal}{Phys. Rev. Lett.} \textbf{\bibinfo{volume}{114}},
  \bibinfo{pages}{170501} (\bibinfo{year}{2015}).

\bibitem[{\citenamefont{Carlini et~al.}(2006)\citenamefont{Carlini, Hosoya,
  Koike, and Okudaira}}]{carlini2006time}
\bibinfo{author}{\bibfnamefont{A.}~\bibnamefont{Carlini}},
  \bibinfo{author}{\bibfnamefont{A.}~\bibnamefont{Hosoya}},
  \bibinfo{author}{\bibfnamefont{T.}~\bibnamefont{Koike}}, \bibnamefont{and}
  \bibinfo{author}{\bibfnamefont{Y.}~\bibnamefont{Okudaira}},
  \emph{\bibinfo{title}{Time-Optimal Quantum Evolution}},
  \bibinfo{journal}{Phys. Rev. Lett.} \textbf{\bibinfo{volume}{96}},
  \bibinfo{pages}{060503} (\bibinfo{year}{2006}).

\bibitem[{\citenamefont{Carlini and Koike}(2012)}]{carlini2012time}
\bibinfo{author}{\bibfnamefont{A.}~\bibnamefont{Carlini}} \bibnamefont{and}
  \bibinfo{author}{\bibfnamefont{T.}~\bibnamefont{Koike}},
  \emph{\bibinfo{title}{Time-Optimal Transfer of Coherence}},
  \bibinfo{journal}{Phys. Rev. A} \textbf{\bibinfo{volume}{86}},
  \bibinfo{pages}{054302} (\bibinfo{year}{2012}).

\bibitem[{\citenamefont{Geng et~al.}(2016)\citenamefont{Geng, Wu, Wang, Xu,
  Shi, Xie, Rong, and Du}}]{geng2016experimental}
\bibinfo{author}{\bibfnamefont{J.}~\bibnamefont{Geng}},
  \bibinfo{author}{\bibfnamefont{Y.}~\bibnamefont{Wu}},
  \bibinfo{author}{\bibfnamefont{X.}~\bibnamefont{Wang}},
  \bibinfo{author}{\bibfnamefont{K.}~\bibnamefont{Xu}},
  \bibinfo{author}{\bibfnamefont{F.}~\bibnamefont{Shi}},
  \bibinfo{author}{\bibfnamefont{Y.}~\bibnamefont{Xie}},
  \bibinfo{author}{\bibfnamefont{X.}~\bibnamefont{Rong}}, \bibnamefont{and}
  \bibinfo{author}{\bibfnamefont{J.}~\bibnamefont{Du}},
  \emph{\bibinfo{title}{Experimental Time-Optimal Universal Control of Spin Qubits in Solids}}, \bibinfo{journal}{Phys. Rev. Lett.}
  \textbf{\bibinfo{volume}{117}}, \bibinfo{pages}{170501}
  (\bibinfo{year}{2016}).

\bibitem[{\citenamefont{Maze et~al.}(2008)\citenamefont{Maze, Stanwix, Hodges,
  Hong, Taylor, Cappellaro, Jiang, Dutt, Togan, Zibrov
  et~al.}}]{maze2008nanoscale}
\bibinfo{author}{\bibfnamefont{J.~R.} \bibnamefont{Maze}},
  \bibinfo{author}{\bibfnamefont{P.~L.} \bibnamefont{Stanwix}},
  \bibinfo{author}{\bibfnamefont{J.~S.} \bibnamefont{Hodges}},
  \bibinfo{author}{\bibfnamefont{S.}~\bibnamefont{Hong}},
  \bibinfo{author}{\bibfnamefont{J.~M.} \bibnamefont{Taylor}},
  \bibinfo{author}{\bibfnamefont{P.}~\bibnamefont{Cappellaro}},
  \bibinfo{author}{\bibfnamefont{L.}~\bibnamefont{Jiang}},
  \bibinfo{author}{\bibfnamefont{M.~G.} \bibnamefont{Dutt}},
  \bibinfo{author}{\bibfnamefont{E.}~\bibnamefont{Togan}},
  \bibinfo{author}{\bibfnamefont{A.}~\bibnamefont{Zibrov}},
  \bibnamefont{et~al.}, \emph{\bibinfo{title}{Nanoscale Magnetic Sensing with an Individual Electronic Spin in Diamond}}, \bibinfo{journal}{Nature}
  \textbf{\bibinfo{volume}{455}}, \bibinfo{pages}{644} (\bibinfo{year}{2008}).

\bibitem[{\citenamefont{Xiang et~al.}(2013)\citenamefont{Xiang, Ashhab, You,
  and Nori}}]{xiang2013hybrid}
\bibinfo{author}{\bibfnamefont{Z.-L.} \bibnamefont{Xiang}},
  \bibinfo{author}{\bibfnamefont{S.}~\bibnamefont{Ashhab}},
  \bibinfo{author}{\bibfnamefont{J.}~\bibnamefont{You}}, \bibnamefont{and}
  \bibinfo{author}{\bibfnamefont{F.}~\bibnamefont{Nori}},
  \emph{\bibinfo{title}{Hybrid Quantum Circuits: Superconducting Circuits Interacting with Other Quantum Systems}}, \bibinfo{journal}{Rev. Mod. Phys.}
  \textbf{\bibinfo{volume}{85}}, \bibinfo{pages}{623} (\bibinfo{year}{2013}).

\bibitem[{\citenamefont{Lu et~al.}(2020)\citenamefont{Lu, Zhang, Liu, Nori,
  Fan, and Pan}}]{lu2020observing}
\bibinfo{author}{\bibfnamefont{Y.-N.} \bibnamefont{Lu}},
  \bibinfo{author}{\bibfnamefont{Y.-R.} \bibnamefont{Zhang}},
  \bibinfo{author}{\bibfnamefont{G.-Q.} \bibnamefont{Liu}},
  \bibinfo{author}{\bibfnamefont{F.}~\bibnamefont{Nori}},
  \bibinfo{author}{\bibfnamefont{H.}~\bibnamefont{Fan}}, \bibnamefont{and}
  \bibinfo{author}{\bibfnamefont{X.-Y.} \bibnamefont{Pan}},
  \emph{\bibinfo{title}{Observing Information Backflow from Controllable Non-Markovian Multichannels in Diamond}}, \bibinfo{journal}{Phys. Rev. Lett.}
  \textbf{\bibinfo{volume}{124}}, \bibinfo{pages}{210502}
  (\bibinfo{year}{2020}).

\bibitem{SM}{See Supplemental Material at [] for the experimental setup, the construction and measurement of single and two-qubit gates, and the algorithm of entanglement-enhanced phase estimation, which includes
Refs. \cite{li2018enhancing,zhao2020improving}.}

\bibitem[{\citenamefont{Li et~al.}(2018)\citenamefont{Li, Dong, Xu, Li, Chen,
  Du, Ge, Guo, and Sun}}]{li2018enhancing}
\bibinfo{author}{\bibfnamefont{C.-H.} \bibnamefont{Li}},
  \bibinfo{author}{\bibfnamefont{Y.}~\bibnamefont{Dong}},
  \bibinfo{author}{\bibfnamefont{J.-Y.} \bibnamefont{Xu}},
  \bibinfo{author}{\bibfnamefont{D.-F.} \bibnamefont{Li}},
  \bibinfo{author}{\bibfnamefont{X.-D.} \bibnamefont{Chen}},
  \bibinfo{author}{\bibfnamefont{A.}~\bibnamefont{Du}},
  \bibinfo{author}{\bibfnamefont{Y.-S.} \bibnamefont{Ge}},
  \bibinfo{author}{\bibfnamefont{G.-C.} \bibnamefont{Guo}}, \bibnamefont{and}
  \bibinfo{author}{\bibfnamefont{F.-W.} \bibnamefont{Sun}},
  \emph{\bibinfo{title}{Enhancing the Sensitivity of a Single Electron Spin Sensor by Multi-Frequency Control}},
  \bibinfo{journal}{Appl. Phys. Lett.} \textbf{\bibinfo{volume}{113}},
  \bibinfo{pages}{072401} (\bibinfo{year}{2018}).

  \bibitem[{\citenamefont{Zhao et~al.}(2020)\citenamefont{Zhao, Dong, Zhang,
  Chen, Zhu, and Sun}}]{zhao2020improving}
\bibinfo{author}{\bibfnamefont{B.}~\bibnamefont{Zhao}},
  \bibinfo{author}{\bibfnamefont{Y.}~\bibnamefont{Dong}},
  \bibinfo{author}{\bibfnamefont{S.}~\bibnamefont{Zhang}},
  \bibinfo{author}{\bibfnamefont{X.}~\bibnamefont{Chen}},
  \bibinfo{author}{\bibfnamefont{W.}~\bibnamefont{Zhu}}, \bibnamefont{and}
  \bibinfo{author}{\bibfnamefont{F.}~\bibnamefont{Sun}},
  \emph{\bibinfo{title}{Improving the NV Generation Efficiency by Electron Irradiation}},
  \bibinfo{journal}{Chin. Opt. Lett.} \textbf{\bibinfo{volume}{18}},
  \bibinfo{pages}{080201} (\bibinfo{year}{2020}).


\bibitem[{\citenamefont{Motzoi et~al.}(2009)\citenamefont{Motzoi, Gambetta,
 Rebentrost, and Wilhelm}}]{motzoi2009simple}
\bibinfo{author}{\bibfnamefont{F.}~\bibnamefont{Motzoi}},
  \bibinfo{author}{\bibfnamefont{J.~M.} \bibnamefont{Gambetta}},
  \bibinfo{author}{\bibfnamefont{P.}~\bibnamefont{Rebentrost}},
  \bibnamefont{and} \bibinfo{author}{\bibfnamefont{F.~K.}
  \bibnamefont{Wilhelm}}, \emph{\bibinfo{title}{Simple Pulses for Elimination of Leakage in Weakly Nonlinear Qubits}}, \bibinfo{journal}{Phys. Rev. Lett.}
  \textbf{\bibinfo{volume}{103}}, \bibinfo{pages}{110501}
  (\bibinfo{year}{2009}).

\bibitem[{\citenamefont{Lloyd}(2000)}]{lloyd2000ultimate}
\bibinfo{author}{\bibfnamefont{S.}~\bibnamefont{Lloyd}},
  \emph{\bibinfo{title}{Ultimate Physical Limits to Computation}},
  \bibinfo{journal}{Nature} \textbf{\bibinfo{volume}{406}},
  \bibinfo{pages}{1047} (\bibinfo{year}{2000}).

\bibitem[{\citenamefont{Lam et~al.}(2021)\citenamefont{Lam, Peter, Groh, Alt,
  Robens, Meschede, Negretti, Montangero, Calarco, and
  Alberti}}]{lam2021demonstration}
\bibinfo{author}{\bibfnamefont{M.~R.} \bibnamefont{Lam}},
  \bibinfo{author}{\bibfnamefont{N.}~\bibnamefont{Peter}},
  \bibinfo{author}{\bibfnamefont{T.}~\bibnamefont{Groh}},
  \bibinfo{author}{\bibfnamefont{W.}~\bibnamefont{Alt}},
  \bibinfo{author}{\bibfnamefont{C.}~\bibnamefont{Robens}},
  \bibinfo{author}{\bibfnamefont{D.}~\bibnamefont{Meschede}},
  \bibinfo{author}{\bibfnamefont{A.}~\bibnamefont{Negretti}},
  \bibinfo{author}{\bibfnamefont{S.}~\bibnamefont{Montangero}},
  \bibinfo{author}{\bibfnamefont{T.}~\bibnamefont{Calarco}}, \bibnamefont{and}
  \bibinfo{author}{\bibfnamefont{A.}~\bibnamefont{Alberti}},
  \emph{\bibinfo{title}{Demonstration of Quantum Brachistochrones between Distant States of an Atom}},
  \bibinfo{journal}{Phys. Rev. X} \textbf{\bibinfo{volume}{11}},
  \bibinfo{pages}{011035} (\bibinfo{year}{2021}).

\bibitem[{\citenamefont{Liu et~al.}(2020)\citenamefont{Liu, Xue, and
  Yung}}]{liu2020brachistochronic}
\bibinfo{author}{\bibfnamefont{B.-J.} \bibnamefont{Liu}},
  \bibinfo{author}{\bibfnamefont{Z.-Y.} \bibnamefont{Xue}}, \bibnamefont{and}
  \bibinfo{author}{\bibfnamefont{M.-H.} \bibnamefont{Yung}},
  \emph{\bibinfo{title}{Brachistochronic Non-Adiabatic Holonomic Quantum Control}}, \bibinfo{journal}{arXiv:2001.05182}  (\bibinfo{year}{2020}).

\bibitem[{\citenamefont{Chen et~al.}(2020)\citenamefont{Chen, Shen, and
  Xue}}]{PhysRevApplied.14.034038}
\bibinfo{author}{\bibfnamefont{T.}~\bibnamefont{Chen}},
  \bibinfo{author}{\bibfnamefont{P.}~\bibnamefont{Shen}}, \bibnamefont{and}
  \bibinfo{author}{\bibfnamefont{Z.-Y.} \bibnamefont{Xue}},
  \emph{\bibinfo{title}{Robust and Fast Holonomic Quantum Gates with Encoding on Superconducting Circuits}}, \bibinfo{journal}{Phys. Rev. Appl.}
  \textbf{\bibinfo{volume}{14}}, \bibinfo{pages}{034038}
  (\bibinfo{year}{2020}).

\bibitem[{\citenamefont{Klei{\ss}ler et~al.}(2018)\citenamefont{Klei{\ss}ler,
  Lazariev, and Arroyo-Camejo}}]{kleissler2018universal}
\bibinfo{author}{\bibfnamefont{F.}~\bibnamefont{Klei{\ss}ler}},
  \bibinfo{author}{\bibfnamefont{A.}~\bibnamefont{Lazariev}}, \bibnamefont{and}
  \bibinfo{author}{\bibfnamefont{S.}~\bibnamefont{Arroyo-Camejo}},
  \emph{\bibinfo{title}{Universal, High-Fidelity Quantum Gates Based on Superadiabatic, Geometric Phases on a Solid-State Spin-Qubit at Room Temperature}}, \bibinfo{journal}{NPJ Quantum Inf.}
  \textbf{\bibinfo{volume}{4}}, \bibinfo{pages}{1} (\bibinfo{year}{2018}).

\bibitem[{\citenamefont{Rong et~al.}(2014)\citenamefont{Rong, Geng, Wang,
  Zhang, Ju, Shi, Duan, and Du}}]{rong2014implementation}
\bibinfo{author}{\bibfnamefont{X.}~\bibnamefont{Rong}},
  \bibinfo{author}{\bibfnamefont{J.}~\bibnamefont{Geng}},
  \bibinfo{author}{\bibfnamefont{Z.}~\bibnamefont{Wang}},
  \bibinfo{author}{\bibfnamefont{Q.}~\bibnamefont{Zhang}},
  \bibinfo{author}{\bibfnamefont{C.}~\bibnamefont{Ju}},
  \bibinfo{author}{\bibfnamefont{F.}~\bibnamefont{Shi}},
  \bibinfo{author}{\bibfnamefont{C.-K.} \bibnamefont{Duan}}, \bibnamefont{and}
  \bibinfo{author}{\bibfnamefont{J.}~\bibnamefont{Du}},
  \emph{\bibinfo{title}{Implementation of Dynamically Corrected Gates on a Single Electron Spin in Diamond}}, \bibinfo{journal}{Phys. Rev. Lett.}
  \textbf{\bibinfo{volume}{112}}, \bibinfo{pages}{050503}
  (\bibinfo{year}{2014}).

\bibitem[{\citenamefont{Chow et~al.}(2012)\citenamefont{Chow, Gambetta,
  Corcoles, Merkel, Smolin, Rigetti, Poletto, Keefe, Rothwell, Rozen
  et~al.}}]{chow2012universal}
\bibinfo{author}{\bibfnamefont{J.~M.} \bibnamefont{Chow}},
  \bibinfo{author}{\bibfnamefont{J.~M.} \bibnamefont{Gambetta}},
  \bibinfo{author}{\bibfnamefont{A.~D.} \bibnamefont{Corcoles}},
  \bibinfo{author}{\bibfnamefont{S.~T.} \bibnamefont{Merkel}},
  \bibinfo{author}{\bibfnamefont{J.~A.} \bibnamefont{Smolin}},
  \bibinfo{author}{\bibfnamefont{C.}~\bibnamefont{Rigetti}},
  \bibinfo{author}{\bibfnamefont{S.}~\bibnamefont{Poletto}},
  \bibinfo{author}{\bibfnamefont{G.~A.} \bibnamefont{Keefe}},
  \bibinfo{author}{\bibfnamefont{M.~B.} \bibnamefont{Rothwell}},
  \bibinfo{author}{\bibfnamefont{J.~R.} \bibnamefont{Rozen}},
  \bibnamefont{et~al.}, \emph{\bibinfo{title}{Universal Quantum Gate Set Approaching Fault-Tolerant Thresholds with Superconducting Qubits}},
  \bibinfo{journal}{Phys. Rev. Lett.} \textbf{\bibinfo{volume}{109}},
  \bibinfo{pages}{060501} (\bibinfo{year}{2012}).

\bibitem[{\citenamefont{Zhang et~al.}(2014)\citenamefont{Zhang, Souza, Brandao,
  and Suter}}]{zhang2014protected}
\bibinfo{author}{\bibfnamefont{J.}~\bibnamefont{Zhang}},
  \bibinfo{author}{\bibfnamefont{A.~M.} \bibnamefont{Souza}},
  \bibinfo{author}{\bibfnamefont{F.~D.} \bibnamefont{Brandao}},
  \bibnamefont{and} \bibinfo{author}{\bibfnamefont{D.}~\bibnamefont{Suter}},
  \emph{\bibinfo{title}{Protected Quantum Computing: Interleaving Gate Operations with Dynamical Decoupling Sequences}}, \bibinfo{journal}{Phys.
  Rev. Lett.} \textbf{\bibinfo{volume}{112}}, \bibinfo{pages}{050502}
  (\bibinfo{year}{2014}).

\bibitem[{\citenamefont{Zhang and Suter}(2015)}]{zhang2015experimental}
\bibinfo{author}{\bibfnamefont{J.}~\bibnamefont{Zhang}} \bibnamefont{and}
  \bibinfo{author}{\bibfnamefont{D.}~\bibnamefont{Suter}},
  \emph{\bibinfo{title}{Experimental Protection of Two-Qubit Quantum Gates against Environmental Noise by Dynamical Decoupling}},
  \bibinfo{journal}{Phys. Rev. Lett.} \textbf{\bibinfo{volume}{115}},
  \bibinfo{pages}{110502} (\bibinfo{year}{2015}).

\bibitem[{\citenamefont{Jelezko et~al.}(2004)\citenamefont{Jelezko, Gaebel,
  Popa, Domhan, Gruber, and Wrachtrup}}]{jelezko2004observation}
\bibinfo{author}{\bibfnamefont{F.}~\bibnamefont{Jelezko}},
  \bibinfo{author}{\bibfnamefont{T.}~\bibnamefont{Gaebel}},
  \bibinfo{author}{\bibfnamefont{I.}~\bibnamefont{Popa}},
  \bibinfo{author}{\bibfnamefont{M.}~\bibnamefont{Domhan}},
  \bibinfo{author}{\bibfnamefont{A.}~\bibnamefont{Gruber}}, \bibnamefont{and}
  \bibinfo{author}{\bibfnamefont{J.}~\bibnamefont{Wrachtrup}},
  \emph{\bibinfo{title}{Observation of Coherent Oscillation of a Single Nuclear Spin and Realization of a Two-Qubit Conditional Quantum Gate}},
  \bibinfo{journal}{Phys. Rev. Lett.} \textbf{\bibinfo{volume}{93}},
  \bibinfo{pages}{130501} (\bibinfo{year}{2004}).

\bibitem[{\citenamefont{Golter et~al.}(2014)\citenamefont{Golter, Baldwin, and
  Wang}}]{golter2014protecting}
\bibinfo{author}{\bibfnamefont{D.~A.} \bibnamefont{Golter}},
  \bibinfo{author}{\bibfnamefont{T.~K.} \bibnamefont{Baldwin}},
  \bibnamefont{and} \bibinfo{author}{\bibfnamefont{H.}~\bibnamefont{Wang}},
  \emph{\bibinfo{title}{Protecting a Solid-State Spin from Decoherence Using Dressed Spin States}}, \bibinfo{journal}{Phys. Rev. Lett.}
  \textbf{\bibinfo{volume}{113}}, \bibinfo{pages}{237601}
  (\bibinfo{year}{2014}).

\bibitem[{\citenamefont{Genov et~al.}(2017)\citenamefont{Genov, Schraft,
  Vitanov, and Halfmann}}]{genov2017arbitrarily}
\bibinfo{author}{\bibfnamefont{G.~T.} \bibnamefont{Genov}},
  \bibinfo{author}{\bibfnamefont{D.}~\bibnamefont{Schraft}},
  \bibinfo{author}{\bibfnamefont{N.~V.} \bibnamefont{Vitanov}},
  \bibnamefont{and} \bibinfo{author}{\bibfnamefont{T.}~\bibnamefont{Halfmann}},
  \emph{\bibinfo{title}{Arbitrarily Accurate Pulse Sequences for Robust Dynamical Decoupling}}, \bibinfo{journal}{Phys. Rev. Lett.}
  \textbf{\bibinfo{volume}{118}}, \bibinfo{pages}{133202}
  (\bibinfo{year}{2017}).

\bibitem[{\citenamefont{Zaiser et~al.}(2016)\citenamefont{Zaiser, Rendler,
  Jakobi, Wolf, Lee, Wagner, Bergholm, Schulte-Herbr{\"u}ggen, Neumann, and
  Wrachtrup}}]{zaiser2016enhancing}
\bibinfo{author}{\bibfnamefont{S.}~\bibnamefont{Zaiser}},
  \bibinfo{author}{\bibfnamefont{T.}~\bibnamefont{Rendler}},
  \bibinfo{author}{\bibfnamefont{I.}~\bibnamefont{Jakobi}},
  \bibinfo{author}{\bibfnamefont{T.}~\bibnamefont{Wolf}},
  \bibinfo{author}{\bibfnamefont{S.-Y.} \bibnamefont{Lee}},
  \bibinfo{author}{\bibfnamefont{S.}~\bibnamefont{Wagner}},
  \bibinfo{author}{\bibfnamefont{V.}~\bibnamefont{Bergholm}},
  \bibinfo{author}{\bibfnamefont{T.}~\bibnamefont{Schulte-Herbr{\"u}ggen}},
  \bibinfo{author}{\bibfnamefont{P.}~\bibnamefont{Neumann}}, \bibnamefont{and}
  \bibinfo{author}{\bibfnamefont{J.}~\bibnamefont{Wrachtrup}},
  \emph{\bibinfo{title}{Enhancing Quantum Sensing Sensitivity by a Quantum Memory}}, \bibinfo{journal}{Nat. Commun.} \textbf{\bibinfo{volume}{7}},
  \bibinfo{pages}{1} (\bibinfo{year}{2016}).

\bibitem[{\citenamefont{Aslam et~al.}(2017)\citenamefont{Aslam, Pfender,
  Neumann, Reuter, Zappe, de~Oliveira, Denisenko, Sumiya, Onoda, Isoya
  et~al.}}]{aslam2017nanoscale}
\bibinfo{author}{\bibfnamefont{N.}~\bibnamefont{Aslam}},
  \bibinfo{author}{\bibfnamefont{M.}~\bibnamefont{Pfender}},
  \bibinfo{author}{\bibfnamefont{P.}~\bibnamefont{Neumann}},
  \bibinfo{author}{\bibfnamefont{R.}~\bibnamefont{Reuter}},
  \bibinfo{author}{\bibfnamefont{A.}~\bibnamefont{Zappe}},
  \bibinfo{author}{\bibfnamefont{F.~F.} \bibnamefont{de~Oliveira}},
  \bibinfo{author}{\bibfnamefont{A.}~\bibnamefont{Denisenko}},
  \bibinfo{author}{\bibfnamefont{H.}~\bibnamefont{Sumiya}},
  \bibinfo{author}{\bibfnamefont{S.}~\bibnamefont{Onoda}},
  \bibinfo{author}{\bibfnamefont{J.}~\bibnamefont{Isoya}},
  \bibnamefont{et~al.}, \emph{\bibinfo{title}{Nanoscale Nuclear Magnetic Resonance with Chemical Resolution}}, \bibinfo{journal}{Science}
  \textbf{\bibinfo{volume}{357}}, \bibinfo{pages}{67} (\bibinfo{year}{2017}).

\end{thebibliography}
%\bibliographystyle{apsrev}

\end{document}

\title{Supplementary Material: Fast high-fidelity geometric quantum control with quantum brachistochrones}
\author{Yang Dong}
\author{Ce Feng}
\author{Yu Zheng}
\author{Xiang-Dong Chen}
\author{Guang-Can Guo}
\author{Fang-Wen Sun}
\email{fwsun@ustc.edu.cn}
\affiliation{{CAS Key Laboratory of Quantum Information, University of Science and Technology of China, Hefei, 230026, People's Republic of China}}
\affiliation{{CAS Center for Excellence in Quantum Information and Quantum Physics, University of Science and Technology of China, Hefei, 230026, People's Republic of China}}
\date{\today }

%\pacs{Valid PACS appear here}
\maketitle

\makeatletter
\renewcommand{\thefigure}{S\@arabic\c@figure}
\makeatother
\makeatletter
\renewcommand{\thetable}{S\@arabic\c@table}
\makeatother
\makeatletter
\renewcommand{\theequation}{S\@arabic\c@equation}
\makeatother

\section{NV center in diamond}
\begin{figure}[bp]
\centering
\textsf{\includegraphics[width=8cm]{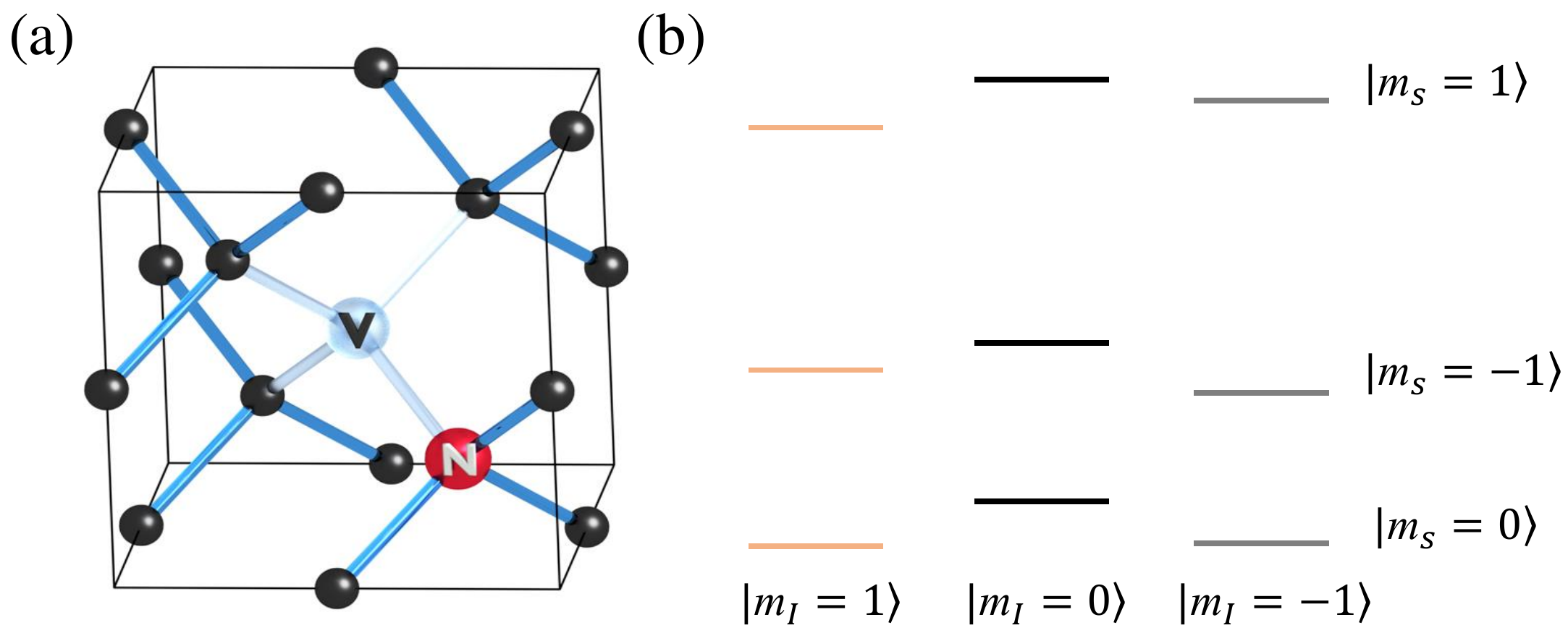}}
\caption{(a) Schematic of the atomic structure of the NV center. (b) Spin energy level diagram of the NV center.}
\label{Sfig1}
\end{figure}

The NV center includes a substitutional nitrogen atom and a vacancy in the nearest-neighbor lattice position as shown in Fig. \ref{Sfig1}(a). In our experiment, a static magnetic field, ${B_0}=510$ G, is applied along the NV symmetry axis ([1 1 1] crystal axis) and removes the degeneracy between the $\left| {{m_S} =   1} \right\rangle $ and $\left| {{m_S} =  - 1} \right\rangle $ electron spin states. Under this magnetic field, the nitrogen nuclear spin can be optically polarized through excited level anticrossing effects \cite{dong2018non,li2018enhancing} and the signal contrast can be enhanced up to ${C = 0.27}$.

The spin energy levels of the NV center are shown in Fig. \ref{Sfig1}(b). We encode $\left| {{m_S} =  - 1} \right\rangle  \equiv \left| 0 \right\rangle $ and $\left| {{m_S} =  1} \right\rangle  \equiv \left| 1 \right\rangle $ ($\left| {{m_I} =  - 1} \right\rangle  \equiv \left| 0 \right\rangle $ and $\left| {{m_I} =  1} \right\rangle  \equiv \left| 1 \right\rangle $) as the qubit basis states and use $\left| {{m_S} = 0} \right\rangle  \equiv \left| a \right\rangle $ ($\left| {{m_I} = 0} \right\rangle  \equiv \left| a \right\rangle $) as an ancillary state for the geometric manipulation of the electron ($^{14}{\text{N}}$ nuclear) spin. The diamond sample is mounted at the focus of a home-built scanning confocal microscope, which is same as described in the Ref. \cite{dong2021quantifying}. After the optical pumping, $98\%$ of the population occupies
the $\left| {{m_I} = 1} \right\rangle $ state of the nuclear spin.
The NV centers studied in this work are formed during chemical vapor deposition growth \cite{zhao2020improving,dong2021quantifying}. The abundance of $^{13}{\text{C}}$ is at the nature level of $1.1\%$.  To beat the fluctuation of the photon counting, we repeat the experimental cycle at least $5 \times {10^7}$ times.

\section{The construction of NHQC gate}
\subsection{Single-qubit gate}
In the single-qubit control case, experiments are implemented on the electron spin qubit while the nuclear spin is kept in state $\left| {{m_I} = 1} \right\rangle $. When MW pulses with the frequencies ${\omega _{1(2)}}$ and  initial phases ${\phi _{1(2)}}$ are applied, the total Hamiltonian of the electron spin qubit is
\begin{eqnarray}
{H_{s}} &=&DS_{z}^{2}+{\gamma _{e}}B{S_{z}}+[{{\gamma _{e}}{B_{1}}\cos
\left( {{\omega _{1}}t+{\phi _{1}}}\right) } \notag\\
&&{+{\gamma _{e}}{B_{2}}\cos \left( {{\omega _{2}}t+{\phi _{2}}}\right) }]{%
S_{x}}\text{.}
\end{eqnarray}

We apply the unitary transformation
\begin{equation}
V=\left( {%
\begin{array}{ccc}
{{e^{-i{\omega _{1}}t}}} & 0 & 0 \\
0 & 1 & 0 \\
0 & 0 & {{e^{-i{\omega _{2}}t}}}%
\end{array}%
}\right)
\end{equation}%
to generate the transformed Hamiltonian ${H_r} = {V^\dag }HV + i\left( {\partial {V^\dag }/\partial t} \right)V$, which is analogous to transforming to the rotating frame in a spin-$1/2$ system. The result is
\begin{equation}
{H_{r}}=\left( {%
\begin{array}{ccc}
{D+{\gamma _{e}}B-{\omega _{2}}} & {\frac{{{\Omega _{2}}}}{2}{e^{-i{\phi _{2}%
}}}} & 0 \\
{\frac{{{\Omega _{2}}}}{2}{e^{i{\phi _{2}}}}} & 0 & {\frac{{{\Omega _{1}}}}{2%
}{e^{i{\phi _{1}}}}} \\
0 & {\frac{{{\Omega _{1}}}}{2}{e^{-i{\phi _{1}}}}} & {D-{\gamma _{e}}B-{%
\omega _{1}}}%
\end{array}%
}\right) \text{,}
\end{equation}%
where ${\Omega _1} = \frac{{\sqrt 2 }}{2}{\gamma _e}{B_1}$ and ${\Omega _2} = \frac{{\sqrt 2 }}{2}{\gamma _e}{B_2}$. By setting ${\omega _1} = D + {\gamma _e}B$ and ${\omega _2} = D - {\gamma _e}B$, we can get

\begin{eqnarray}
{H} &=&\left[ {\frac{{{\Omega _{1}}(t)}}{2}{e^{i{\phi _{1}}}}|0\rangle {%
\text{ + }}\frac{{{\Omega _{2}}(t)}}{2}{e^{i{\phi _{2}}}}|1\rangle }\right]
\left\langle {a}\right\vert +H.c.  \notag \\
&=&\frac{{\Omega (t)}}{2}{e^{i\phi _{2}}\left\vert b\right\rangle }\langle
a|+H.c. \text{,}  \label{EQ1}
\end{eqnarray}%
where ${\left\vert b\right\rangle ={e^{i\phi }}\sin \frac{\theta }{2}%
|0\rangle {\text{ + cos}}\frac{\theta }{2}|1\rangle }$ is a bright state with $\tan \frac{\theta }{2}=\frac{{\Omega
_{1}}}{{\Omega _{2}}}$, 
%Here, ${\Omega _{1}}$ and ${\Omega _{2}}$ are the time-dependent amplitudes of the two microwaves. 
$\phi ={\phi _{1}}-{\phi _{2}}$, and $\Omega (t)=\sqrt{\Omega
_{1}^{2}(t)+\Omega _{2}^{2}(t)}$.
As shown in Eq. (\ref{EQ1}), the bright state ${\left\vert
b\right\rangle }$ interacts with the state $\left\vert a\right\rangle $, while decouples from the dark state ${\left| d \right\rangle  = \cos \frac{\theta }{2}|0\rangle  - {e^{ - i\phi }}{\text{ sin}}\frac{\theta }{2}|1\rangle }$.

\begin{figure}[tbp]
\centering
\textsf{\includegraphics[width=7cm]{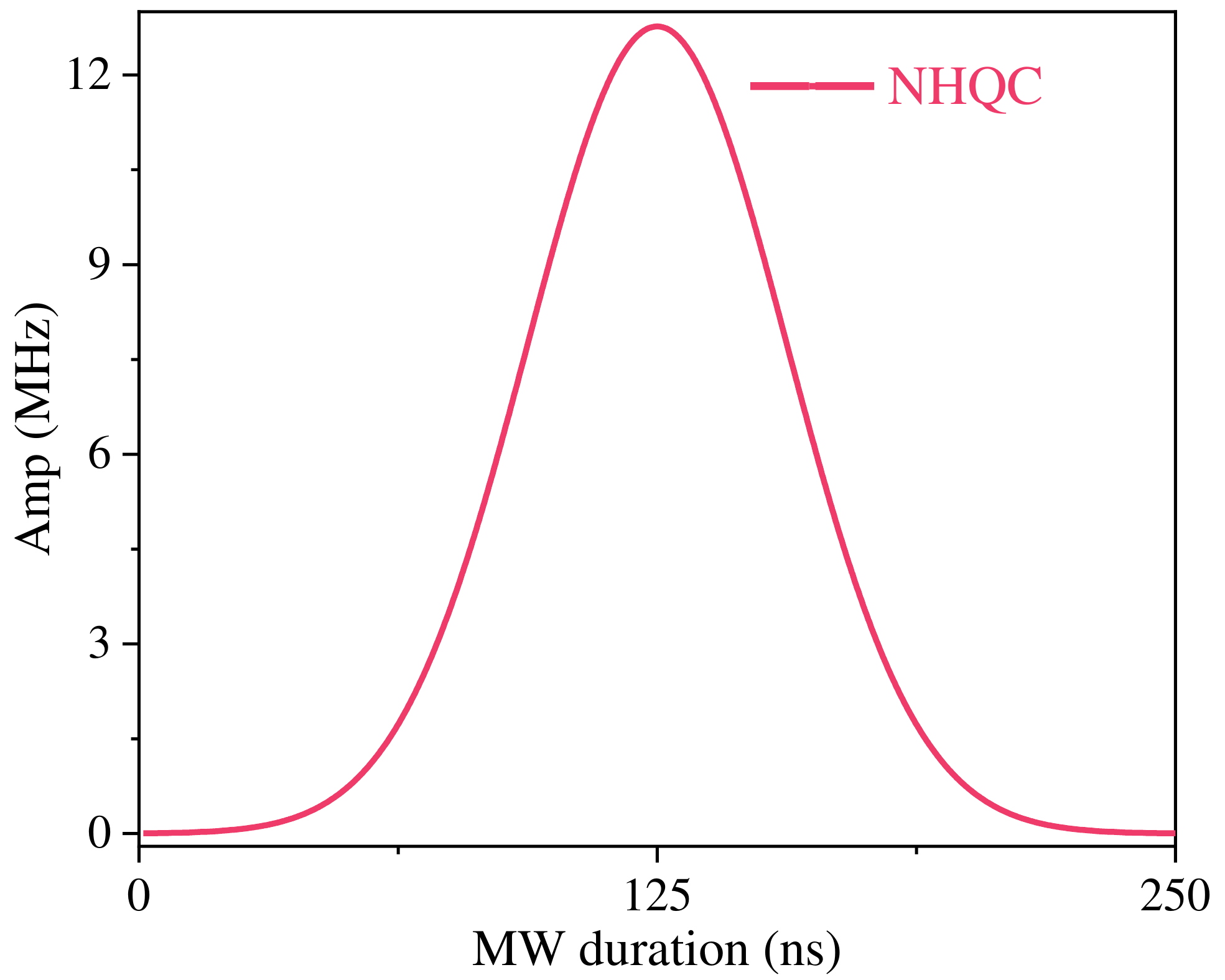}}
\caption{The waveform of the $4\sigma $ truncated Gaussian shape for the conventional NHQC gate.}
\label{Sfig2}
\end{figure}

The NHQC scheme \cite{dong2021experimental} can be realized with a single-loop scenario by engineering the quantum system to evolve along with a red-slice-shaped path in the Bloch sphere, as shown in Fig. 1(a) in the main manuscript. In the first segment $\left[ {0,\tau/2} \right]$, we set ${\phi _2} = 0$, and then ${H}$ is reduced to ${{H}_a} = \frac{\Omega }{2}\left( {\left| b \right\rangle \left\langle a \right| + \left| a \right\rangle \left\langle b \right|} \right)$. The corresponding evolution operator is ${U_a} = \left| d \right\rangle \left\langle d \right| - i\left( {\left| b \right\rangle \left\langle a \right| + \left| a \right\rangle \left\langle b \right|} \right)$. In the second segment $\left[ {\tau/2,\tau} \right]$, we change the MW phase ${\phi _2}$ to ${{\tilde \phi }_2} = \pi  + \gamma $. Then ${H}$ is reduced to ${H_b} =  - \frac{\Omega }{2}\left( {{e^{i\gamma }}|b\rangle \left\langle a \right| + {e^{ - i\gamma }}\left| a \right\rangle \left\langle b \right|} \right)$ and the corresponding evolution operator is ${U_b} = \left| d \right\rangle \langle d| + i\left( {{e^{i\gamma }}|b\rangle \left\langle a \right| + {e^{ - i\gamma }}\left| a \right\rangle \left\langle b \right|} \right)$. In this way, in the qubit computational basis $\left\{ {|0\rangle ,|1\rangle } \right\}$, the induced gate operation will be
\begin{equation}
{U_G}(\gamma ,\theta ,\phi ) = {U_b}{U_a} = {e^{i(\gamma /2)}}{e^{ - i(\gamma /2)\vec n \cdot \vec \sigma }}\text{,}
\label{EQ2}
\end{equation}
which describes a rotation around the $\vec n=(\sin \theta \cos \phi, \sin \theta \sin \phi, \cos \theta)$ axis by a $\gamma$ angle.

In conventional NHQC construction, the cyclic path evolution area is set as $\int_{0}^{\tau} \Omega(t) d t=2\pi$, with $\tau$ is the entire evolution time, which is separated into two equal segments. The geometric control of the $^{14}{\text{N}}$ nuclear spin is same with the single-qubit gate due to similar energy levels. For realistic quantum control of the physical system, the envelopes of the two MWs are Gaussian $4\sigma $ truncated to eliminate unwanted population leakage effect as shown in Fig. \ref{Sfig2} \cite{motzoi2009simple,abdumalikov2013experimental}. The maximum of Rabi frequency with the Gaussian $4\sigma $ truncated pulse is $12.76$ MHz for electron spin operation of the NV center.

\subsection{Two-qubit gate}
In the construction of two-qubit gate, we use the electron spin state as the target state and the non-zero host nitrogen nuclear spin ($I{\text{ = }}1$ for $^{14}\text{N}$) of the NV center as the control qubit. The Hamiltonian can be expressed as:
\begin{equation}
{H_s} = DS_z^2 + {\gamma _e}B{S_z} + PI_z^2 + {\gamma _n}B{I_z} + A_{zz}{S_z}{I_z} \text{.}
\label{H}
\end{equation}
Here, $S_{z}$ and $I_{z}$ are the $z$ components of the spin-$1$ operators for the electron and nitrogen nuclear spins ($^{14}{\text{N}}$), respectively.
${\gamma _e}$ (${\gamma _N}$) is the electronic ($^{14}{\text{N}}$ nuclear) gyromagnetic ratio, ${S_z}$ and ${I_z}$ are the electron and nitrogen nuclear spin operators, respectively. The zero field splitting ${D} = 2.87$ GHz, the nuclear quadrupolar splitting $P =  - 4.95$ MHz, and the
hyperfine interaction between the NV electron and the $^{14}{\text{N}}$ nuclear spin $A = 2.16$ MHz.
By applying state-selective MW and radio-frequency (RF) pulses, we can couple different energy levels. For arbitrary sublevels of the electron spin ground state, the $^{14}{\text{N}}$ nuclear spin system has a lambda ($\Lambda $) energy structure, which is preferable to realize the B-NHQC scheme \cite{liu2019plug,li2020fast}.
%Similarly, we encode $\left| {{m_I} =  - 1} \right\rangle  \equiv \left| 0 \right\rangle $ and $\left| {{m_I} =  1} \right\rangle  \equiv \left| 1 \right\rangle $ as the qubit basis states and use $\left| {{m_I} = 0} \right\rangle  \equiv \left| a \right\rangle $ as an ancillary state for the geometric manipulation of the $^{14}{\text{N}}$ nuclear spin. For this type of the solid hybrid spin system, the two-qubit gate poses some challenges in the experiment, mostly because the characteristic properties of the two types of the spins differ by three orders of magnitude.
\subsection{Quantum process tomography (QPT) of single- and two-qubit gate}
\begin{figure*}[bp]
\centering
\textsf{\includegraphics[width=14cm]{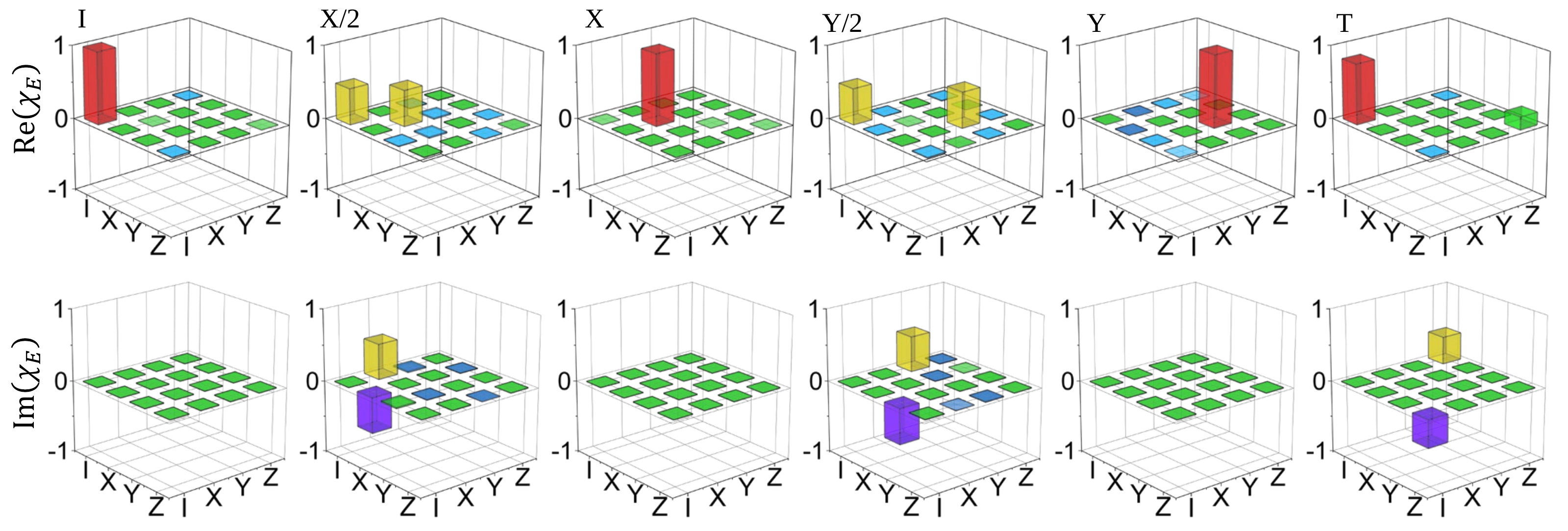}}
\caption{QPT of the B-NHQC gates. A set of the rotations $\left\{ {I,X/2,X,Y/2,Y,T} \right\}$ is used to generate input states and analyze the process. The bar charts of the real and imaginary parts of ${\chi _E}$ for specific gates $\left\{ {I,X/2,X,Y/2,Y,T} \right\}$, giving an average process fidelity of ${F_{av}} = 0.984(2)$. The labels in the $X$ and $Y$ axes correspond to the operators in the basis set $\left\{ {I,{\sigma _x},{\sigma _y},{\sigma _z}} \right\}$ of the $\left| 0 \right\rangle ,\left| 1 \right\rangle $ subspace. }
\label{fig2}
\end{figure*}

We characterize the single-qubit B-NHQC gates through a standard quantum process tomography (QPT) method \cite{dong2021quantifying}. The experimental process matrices ${\chi _E}$ of six specific geometric gates $\left\{ {I,X/2,X,Y/2,Y,T} \right\}$ are shown in Fig. \ref{fig2}, with the fidelities of 0.988(4), 0.981(3), 0.982(3), 0.981(6), 0.987(7), 0.983(4), respectively. Here, the quantum process fidelity is calculated with $F = | {Tr( {{\chi _E}\chi _{id}^\dag })}|$ by comparing with the idea operation ${\chi _{id}}$. The major contribution to the QPT infidelity comes from state preparation and detection errors \cite{zu2014experimental,arroyo2014room,kleissler2018universal,huang2019experimental}.

\begin{figure*}[tbp]
\centering
\textsf{\includegraphics[width=10cm]{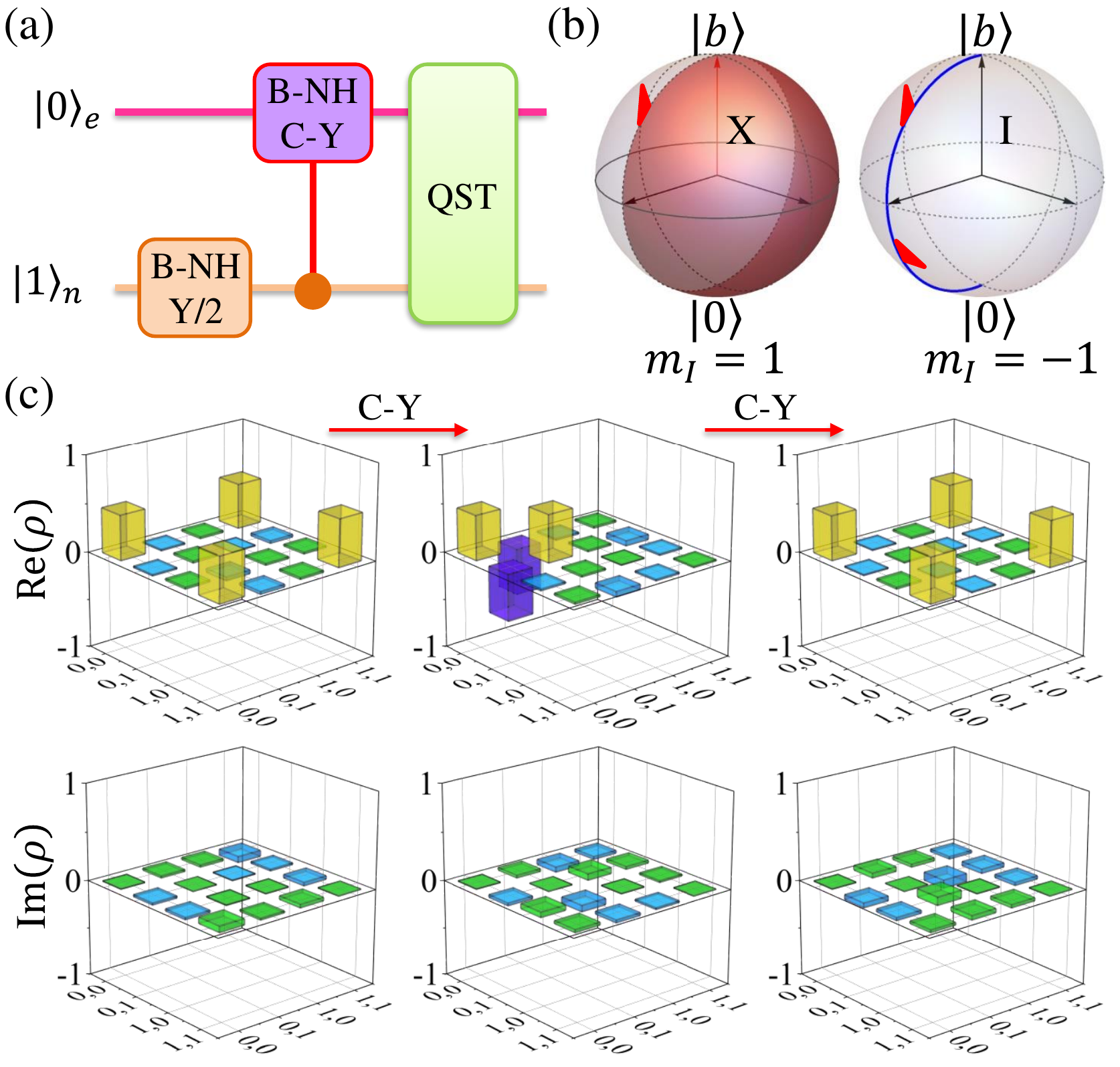}}
\caption{(a) The quantum circuit to evaluate the B-NHQC two-qubit gate. (b) Dynamical process of the B-NHQC C-Y gate with Bloch sphere. (c) The density matrices of the two-qubit states are reconstructed by the two-qubit QST to show the transformation by the B-NHQC C-Y gate. The bars indicate the values of the matrix elements.}
\label{figs4}
\end{figure*}

Fig. \ref{figs4}(a) shows the quantum circuit to evaluate the B-NHQC two-qubit gate. The dynamical process of the B-NHQC C-Y gate with Bloch sphere is shown in Fig.\ref{figs4}(b). Fig.\ref{figs4}(c) demonstrates the density matrices of the two-qubit states to show the transformation by the B-NHQC C-Y gate, which are reconstructed by the two-qubit quantum state tomography (QST).

\section{Entanglement-enhanced phase estimation algorithm }

Fig. 3(a) in the main manuscript shows the quantum circuit to prepare the quantum states and probe the phase \cite{giovannetti2004quantum,giovannetti2006quantum,pirandola2018advances,pezze2018quantum,dong2016reviving,liu2015demonstration}. For conventional phase estimation procedure, the quantum sensing qubit is transformed to a superposition state $\left( {\left| 0 \right\rangle  + \left| 1 \right\rangle } \right)/\sqrt 2 $ by a $Y/2$ gate. After applying phase gate ${Z_\varphi }$, $\left| 0 \right\rangle $ and $\left| 1 \right\rangle $ acquire a relative phase $\varphi $ and the sensor qubit evolves to $\left( {\left| 0 \right\rangle  + {e^{ - i\varphi }}\left| 1 \right\rangle } \right)/\sqrt 2 $, where the phase $\varphi $ can encode the quantity to be measured. The interferometer signal is the expectation value $P = ({{1 + \cos \varphi }})/{2}$. If we repeat the same interferometric procedure with $2$ uncorrelated sensor qubits, the signal will be simply given by $2$ times that of a single sensor qubit and the standard deviation of the photon-shot-noise will decrease by $\sqrt 2 $, showing the standard quantum limit. We equally select several values of $\varphi $ in $\left[ {0,\pi } \right]$ and implement the corresponding phase gate with NV center. The signal intensity has cosine dependence on the phase of input state, while the measurement photon-shot-noise at the output is denoted with error bars. Moreover, the electron and nuclear spin can be prepared in an entangled state by a combination of the MW and radio-frequency pulses. We implement a phase gate on both sensor qubits to bring the system to the two-qubit NOON state
$\left( {\left| {00} \right\rangle  + {e^{{-}2{\text{i}}\varphi }}\left| {11} \right\rangle } \right)/\sqrt 2 $. The interferometer signal from entangled sensors is $P = ({{1 + \cos 2\varphi }})/{2}$. The phase relation of the entangled state has double frequency dependence on the phase, which means the phase estimation employing the NOON state is more sensitive than that of two uncorrelated sensor qubits.

\begin{table}[tbp]
\centering
\caption{Summary of the experimental parameters for entanglement-enhanced phase estimation algorithm.}
\tabcolsep0.1in
\begin{threeparttable}
\begin{tabular}{c c c c c c}
  \hline\hline
  Term & ${T_{ini}}$($\mu$s)$^1$ & $T_a$($\mu$s)$^2$ & ${T_r}$($\mu$s)$^3$ & $\sigma _S^n$ & Visibility  \\
  \hline
  NHQC & 3 & 287 & 2 & 0.044(4) & 0.90(5)  \\
  B-NHQC & 3 & 80 & 2 & 0.031(3) & 0.97(1)  \\
  \hline\hline
\end{tabular}
      \begin{tablenotes}
        \footnotesize
        \item[1] ${T_{ini}}$ denotes initialization time of the sensing qubits.
        \item[2] $T_a$ denotes duration time of the entanglement-enhanced phase estimation algorithm as shown in Fig. 3(d) in the main manuscript.
        \item[3] ${T_r}$ denotes readout time of the sensing qubits.
      \end{tablenotes}
    \end{threeparttable}
\end{table}